\documentclass[aps,preprint,amsmath,amssymb,t1,11pt]{revtex4}
\usepackage{graphicx}
\usepackage{epstopdf}
\usepackage{xcolor}
\usepackage{slashed}
\usepackage{multirow}
\usepackage{booktabs} 
\usepackage{adjustbox}
\usepackage{rotating}
\usepackage{float}
\usepackage{placeins}
\usepackage{siunitx}
\usepackage{tabularx}
\begin{document}

\title{Sensitivity to top-quark FCNC interactions at future muon colliders}

\author{A. Senol}
\email[]{senol_a@ibu.edu.tr}
\affiliation{Department of Physics, Bolu Abant Izzet Baysal University, 14280, Bolu, T\"{u}rkiye.}
\author{B. S. Ozaltay}
\email[]{sahlanbaris@gmail.com}
\affiliation{Department of Physics, Bolu Abant Izzet Baysal University, 14280, Bolu, T\"{u}rkiye.}
\author{M. Tekin}
\email[]{wustafaatekin@gmail.com}
\affiliation{Department of Physics, Bolu Abant Izzet Baysal University, 14280, Bolu, T\"{u}rkiye.}
\author{H. Denizli}
\email[]{denizli_h@ibu.edu.tr}
\affiliation{Department of Physics, Bolu Abant Izzet Baysal University, 14280, Bolu, T\"{u}rkiye.}

\date{\today}

\begin{abstract}
We investigate flavor-changing neutral current (FCNC) interactions of the top quark at a future muon collider operating at a center-of-mass energy of $\sqrt{s}=10~\mathrm{TeV}$. The process $\mu^{+}\mu^{-}\rightarrow\nu_{\mu}\,\mu^{+}\,b\,j$ together with its charge-conjugate channel is considered as a probe of anomalous $tqZ$ and $tq\gamma$ interactions within a model-independent effective field theory framework. Starting from the most general FCNC Lagrangian, we examine the contributions of vector and tensor operators as well as the chiral structure of the anomalous interactions. Owing to the strong chiral suppression of the right-handed contributions at multi-TeV energies, the detector-level sensitivity analysis is performed for the left-handed tensor couplings. Signal and Standard Model background events are generated using a complete Monte Carlo simulation chain, including matrix-element generation, parton showering and hadronization with \texttt{Pythia}, and fast detector simulation with \texttt{Delphes} using a dedicated $10$~TeV muon collider detector card. Signal discrimination is optimized through a boosted decision tree (BDT) analysis employing an extended set of kinematic observables. Assuming an integrated luminosity of $10~\mathrm{ab}^{-1}$, we derive projected exclusion and discovery sensitivities from a simultaneous two-dimensional scan of the real and imaginary components of the anomalous left-handed couplings $\kappa_{qt}^{L}$ and $\lambda_{qt}^{L}$. The obtained limits reach the $\mathcal{O}(10^{-3})$ level for the effective couplings, corresponding to branching-ratio sensitivities of $\mathcal{O}(10^{-6})$ for the rare decays $t\rightarrow qZ$ and $t\rightarrow q\gamma$. These projections improve the current experimental limits from the ATLAS and CMS collaborations by approximately one order of magnitude.
\end{abstract}
\maketitle

\section{Introduction}
The investigation of rare top-quark processes constitutes a key avenue for probing physics beyond the Standard Model. Owing to its exceptionally large mass and its direct coupling to the electroweak symmetry breaking sector, the top quark occupies a unique position among Standard Model particles. In particular, flavor-changing neutral current (FCNC) interactions involving the top quark are of special interest, as they are strongly suppressed within the Standard Model (SM) and thus provide a clean environment to search for non-standard effects.

Within the SM, FCNC transitions such as $t \to q Z$ and $t \to q \gamma$ ($q = u, c$) arise only at higher orders and are further reduced by the Glashow--Iliopoulos--Maiani mechanism \cite{Glashow70}. As a result, their predicted rates are highly suppressed, rendering them experimentally inaccessible at current collider facilities. This extreme suppression implies that any measurable deviation from the SM expectation would necessarily point to the presence of new degrees of freedom or modified interaction structures. 
Therefore, the possibility of a detectable deviation of FCNC couplings of the top quark from SM predictions has been studied in several BSM models, such as the two Higgs pair model \cite{Eilam:1990zc}, supersymmetry \cite{Yang:1997dk}, technicolor \cite{Lu:2003yr}, and minimal supersymmetric standard model \cite{Li:1993mg}. In these models, the branching ratios of top-quark FCNC decays are significantly enhanced compared to the Standard Model predictions, reaching values of the order of $\mathcal{O}(10^{-6} \text{--} 10^{-5})$. A model-independent manner through higher-dimensional operators within an effective field theory (EFT) description \cite{AguilarSaavedra:2004wm,AguilarSaavedra:2008zc,Aguilar-Saavedra:2009ygx} can significantly enhance FCNC-induced top-quark processes, making them accessible at future high-energy colliders.

First experimental studies on the determination of the experimental limits of the branching ratio of FCNC $t\to q Z$ and $t\to q \gamma$  decays started at the Large Electron-Positron Collider (LEP) \cite{OPAL:2001spi,ALEPH:2002wad,L3:2002hbp,DELPHI:2003cnx}, followed at the Hadron-Electron Ring Accelerator (HERA) \cite{ZEUS:2011mya}, and Tevatron \cite{CDF:2008mpz,D0:2011pcl}, and continue currently at the Large Hadron Collider (LHC) with the ATLAS and CMS collaborations \cite{ATLAS:2015vhj,ATLAS:2018zsq,ATLAS:2022per, ATLAS:2023qzr, CMS:2013knb,CMS:2017wcz,CMS:2023bjm}. 
Among these studies, the most stringent current experimental upper limits at the 95\% confidence level (C.L.) on the branching ratios of the decays $t \to u(c) Z$ have been obtained at the Large Hadron Collider. Results from the CMS Collaboration, based on proton--proton collision data at $\sqrt{s} = 8$ TeV with an integrated luminosity of 19.7 fb$^{-1}$, reported upper limits of ${\rm BR}(t \to uZ) < 2.2 \times 10^{-4}$ and ${\rm BR}(t \to cZ) < 4.9 \times 10^{-4}$ at the 95\% C.L. \cite{CMS:2017wcz}. Furthermore, using proton--proton collision data at $\sqrt{s} = 13$ TeV with an integrated luminosity of 139 fb$^{-1}$, the ATLAS Collaboration has set upper limits of ${\rm BR}(t \to uZ) < 6.2 \times 10^{-5}$ and ${\rm BR}(t \to cZ) < 1.2 \times 10^{-4}$ \cite{ATLAS:2023qzr}. Additionally, the ATLAS (and CMS) Collaboration, using proton-proton collision data at $\sqrt{s} = 13$ TeV with an integrated luminosity of 139 fb$^{-1}$, has determined the upper limits for ${\rm BR}(t \to u\gamma)$ and ${\rm BR}(t \to c\gamma) $ at 95\% C.L. as $0.85 \times 10^{-5}$ ($0.95 \times 10^{-5}$) and $4.4 \times 10^{-5}$ ($1.51 \times 10^{-5}$)  ~\cite{ATLAS:2022per,CMS:2023bjm}. 

Nevertheless, the hadronic initial state inherently limits the achievable sensitivity due to substantial backgrounds and complex event topologies. Complementary to these efforts, lepton collider scenarios have been extensively explored at the phenomenological level, primarily motivated by their clean experimental environment and the absence of strong interactions in the initial state. In this context, the sensitivity to anomalous top-quark FCNC couplings has been widely investigated within effective field theory frameworks at future $e^+e^-$ ~\cite{Han:1998yr,Yang:2004af,Khanpour:2014xla,Tizchang:2024ctw,Khatibi:2021phr,Shi:2019epw,CLICdp:2018esa} and $ep$ \cite{Denizli:2017cfx,TurkCakir:2017rvu,Cakir:2018ruj} colliders

Among the proposed lepton collider options, muon colliders occupy a particularly distinctive position. Such facilities were first emphasized in the 2020 European Strategy for Particle Physics~\cite{mu1,mu2,mu3}. Owing to the large mass of the muon, synchrotron radiation effects are significantly suppressed, allowing circular machines to reach multi-TeV center-of-mass energies while maintaining high luminosity. In this context, benchmark scenarios with $\sqrt{s}=10~\mathrm{TeV}$ and integrated luminosities of the order of $10~\mathrm{ab}^{-1}$ have been widely discussed, offering an excellent opportunity to explore both precision measurements and rare processes. The unique combination of high energy reach and a relatively clean experimental environment makes muon colliders especially well suited for detailed studies in the top-quark sector.

At multi-TeV energies, muon collisions are characterized by a rich interplay of production mechanisms. While electroweak vector boson fusion (VBF) processes dominate the total cross section at high energies~\cite{mu6,mu7}, direct $\mu^+\mu^-$ annihilation channels, such as top-quark pair production and Higgs-associated processes, remain experimentally valuable due to their distinctive kinematic signatures. In particular, final states originating from direct annihilation exhibit invariant mass distributions sharply peaked around $\sqrt{s}$, in contrast to VBF-induced processes where the invariant mass typically remains well below the collider energy. This complementarity enhances the overall sensitivity to both Standard Model parameters and possible new physics effects.

Within this framework, top-quark physics constitutes a central component of the muon collider program. The large mass of the top quark and its strong coupling to the electroweak sector render it a particularly sensitive probe of physics beyond the Standard Model. Moreover, the clean environment of a muon collider enables precise reconstruction of top-quark final states, leading to improved background suppression and reduced systematic uncertainties. Phenomenological studies have demonstrated that muon colliders can achieve significant sensitivity to anomalous top-quark couplings within effective field theory frameworks, including FCNC interactions probed via single-top and associated production channels~\cite{Ake:2023xcz,Han:2024gan,Accettura:2023ked}.

Therefore, muon colliders provide a powerful and complementary platform for advancing top-quark FCNC searches, combining high energy, large event samples, and clean experimental conditions, and thereby offering sensitivity beyond the reach of current hadron collider experiments.

In this study, we investigate top-quark FCNC interactions with the $Z$ boson and the photon by examining the process
\[
\mu^{+}\mu^{-} \to \nu_{\mu}\,\mu^+\,b\,j ,
\]
together with its charge-conjugate channel. This process provides direct sensitivity to anomalous $tqZ$ and $tq\gamma$ vertices through single top-quark production mechanisms. Unless explicitly stated otherwise, all results presented in this work include the contributions from both the signal process and its corresponding charge-conjugate channel. The resulting final-state topology, characterized by a charged muon, missing transverse energy associated with the neutrino, a $b$-tagged jet, and a light-flavor jet, offers a distinctive experimental signature that can be efficiently isolated in the clean environment of a muon collider.

The analysis is carried out within an effective field theory framework, enabling a systematic parametrization of potential deviations from the Standard Model predictions. Focusing on a benchmark muon collider configuration with a center-of-mass energy of $\sqrt{s}=10~\mathrm{TeV}$ and an integrated luminosity of $10~\mathrm{ab}^{-1}$, we derive projected sensitivities to the anomalous FCNC couplings. The results demonstrate that the process $\mu^{+}\mu^{-} \to \nu_{\mu}\,\mu^+\,b\,j$ constitutes a powerful and complementary probe of top-quark FCNC interactions, highlighting the strong discovery potential of future muon collider facilities for uncovering indirect signatures of physics beyond the Standard Model.

The remainder of this paper is organized as follows. In Section~\ref{section2}, we introduce the effective field theory framework and describe the anomalous FCNC $tqZ$ and $tq\gamma$ couplings considered in this study. Section~\ref{section3} is devoted to the generation of signal and background events, together with the implementation of the event selection strategy. In Section~\ref{section4}, we present the details of the multivariate analysis based on boosted decision trees and discuss its impact on the signal--background discrimination. The projected sensitivities and limits on the anomalous $tqZ$ and $tq\gamma$ couplings are reported in Section~\ref{section5}. Finally, the main results and conclusions of the study are summarized in Section~\ref{section6}.

\section{The anomalous FCNC $tq\gamma$ and $tqZ$ couplings}\label{section2}

Within the framework of effective field theory (EFT), flavor-changing neutral current (FCNC) interactions of the top quark are described by higher-dimensional operators added to the renormalizable Standard Model (SM) Lagrangian \cite{AguilarSaavedra:2004wm,AguilarSaavedra:2008zc,Aguilar-Saavedra:2009ygx}. In this work, we adopt the most general effective FCNC Lagrangian describing the $tq\gamma$ and $tqZ$ interactions following the formalism of Ref.~\cite{AguilarSaavedra:2008zc}, which is given by
\begin{eqnarray}\label{eq1}
\mathcal{L}_{eff} & = & -\frac{e}{2 m_t} \bar q   \sigma^{\mu \nu} \left( \lambda_{qt}^L P_L + \lambda_{qt}^R P_R \right) t A_{\mu\nu} - \frac{g}{2 \cos\theta_W} \bar q  \gamma^{\mu} \left( X_{qt}^L P_L + X_{qt}^R P_R \right) t Z_{\mu}\nonumber\\ &&+ \frac{g}{2 \cos\theta_W} \bar q  \frac{ \sigma^{\mu \nu}}{M_Z} \left( \kappa_{qt}^L P_L + \kappa_{qt}^R P_R \right) t Z_{\mu\nu} +\text{H.c.}  
\end{eqnarray}
where $e$ and $g$ denote the electromagnetic and weak gauge couplings, respectively, $\theta_W$ is the Weinberg angle, and $P_{L,R}=(1\mp\gamma_5)/2$ are the left- and right-handed chirality projection operators. The coefficients $X_{qt}^{L,R}$ parameterize the vector-type ($\gamma^\mu$) FCNC interaction between the top quark, a light up-type quark ($q=u,c$), and the $Z$ boson. The coefficients $\kappa_{qt}^{L,R}$ describe the corresponding anomalous tensor ($\sigma^{\mu\nu}$) interaction with the $Z$ boson, while $\lambda_{qt}^{L,R}$ parameterize the  the tensor interaction of the top quark with the photon. In general, all anomalous couplings are complex parameters, whose real and imaginary components may contribute to the phenomenology of FCNC interactions. The field strength tensors are defined as $A_{\mu\nu}$ and $Z_{\mu\nu}$ for the photon and $Z$ boson, respectively. and the antisymmetric tensor is defined as $\sigma^{\mu\nu}=i[\gamma^\mu,\gamma^\nu]/2$.

It should be noted that, within the complete Standard Model Effective Field Theory (SMEFT), the process studied here may also receive contributions from dimension-six four-fermion operators in addition to the anomalous neutral-current FCNC interactions considered in Eq.~(1)~\cite{Aguilar-Saavedra:2010uur,Grzadkowski:2010es,Aguilar-Saavedra:2018ksv}. Such contact interactions represent an independent class of effective operators and may contribute to single-top production depending on the underlying ultraviolet completion. In the present work, however, we deliberately restrict our analysis to anomalous $tqZ$ and $tq\gamma$ interactions in order to provide a dedicated assessment of the sensitivity of a future high-energy muon collider to these FCNC vertices. A global SMEFT analysis including both FCNC and four-fermion operators is beyond the scope of this study and is left for future investigation.

\begin{figure}[htbp]
    \centering
    \includegraphics[scale=0.4]{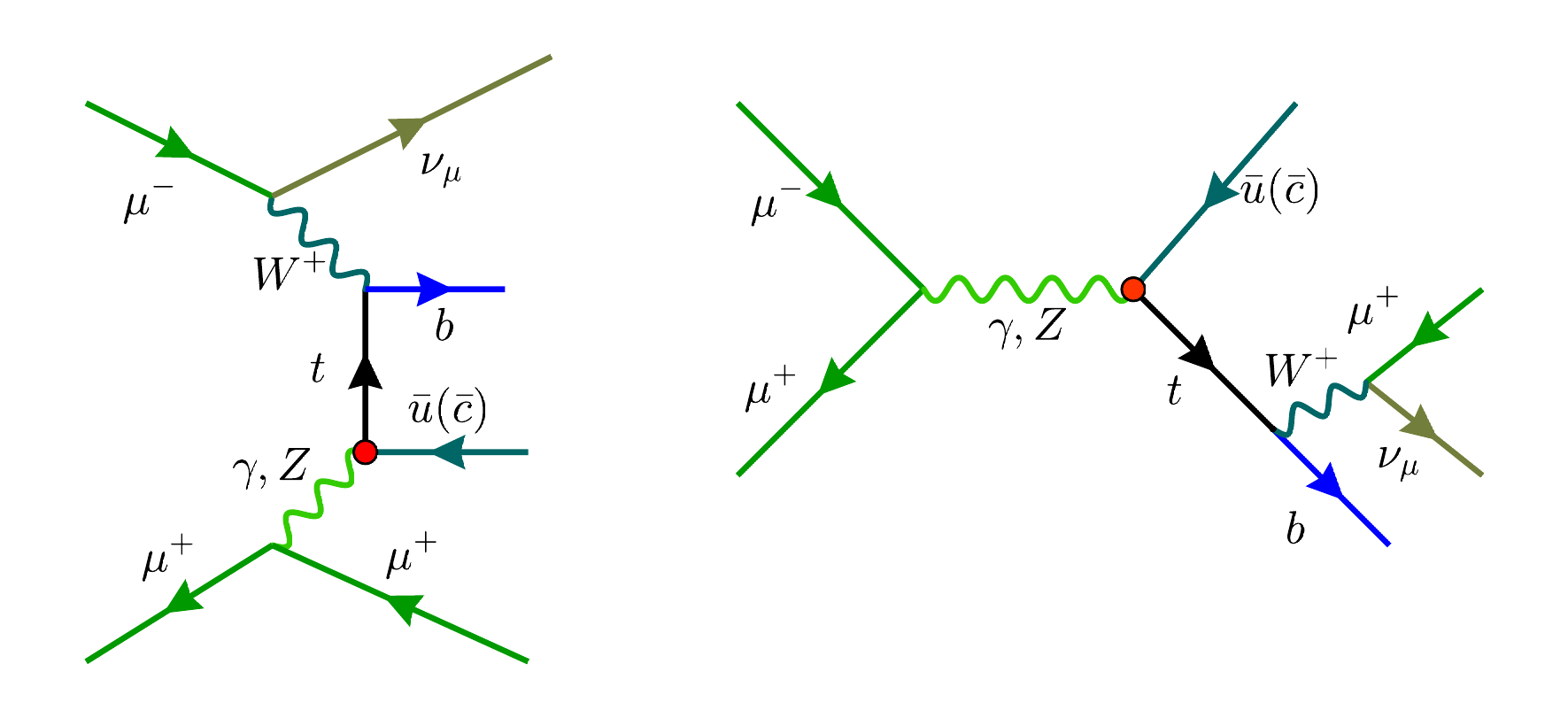}
   \caption{Representative Feynman diagrams for the signal process $\mu^{+}\mu^{-} \to \nu_{\mu}\,\mu^{+}\,b\,j$ induced by anomalous FCNC $tqZ$ and $tq\gamma$ couplings.}
    \label{fd}
\end{figure}

\begin{figure}[htbp]
    \centering
     \includegraphics[scale=0.32]{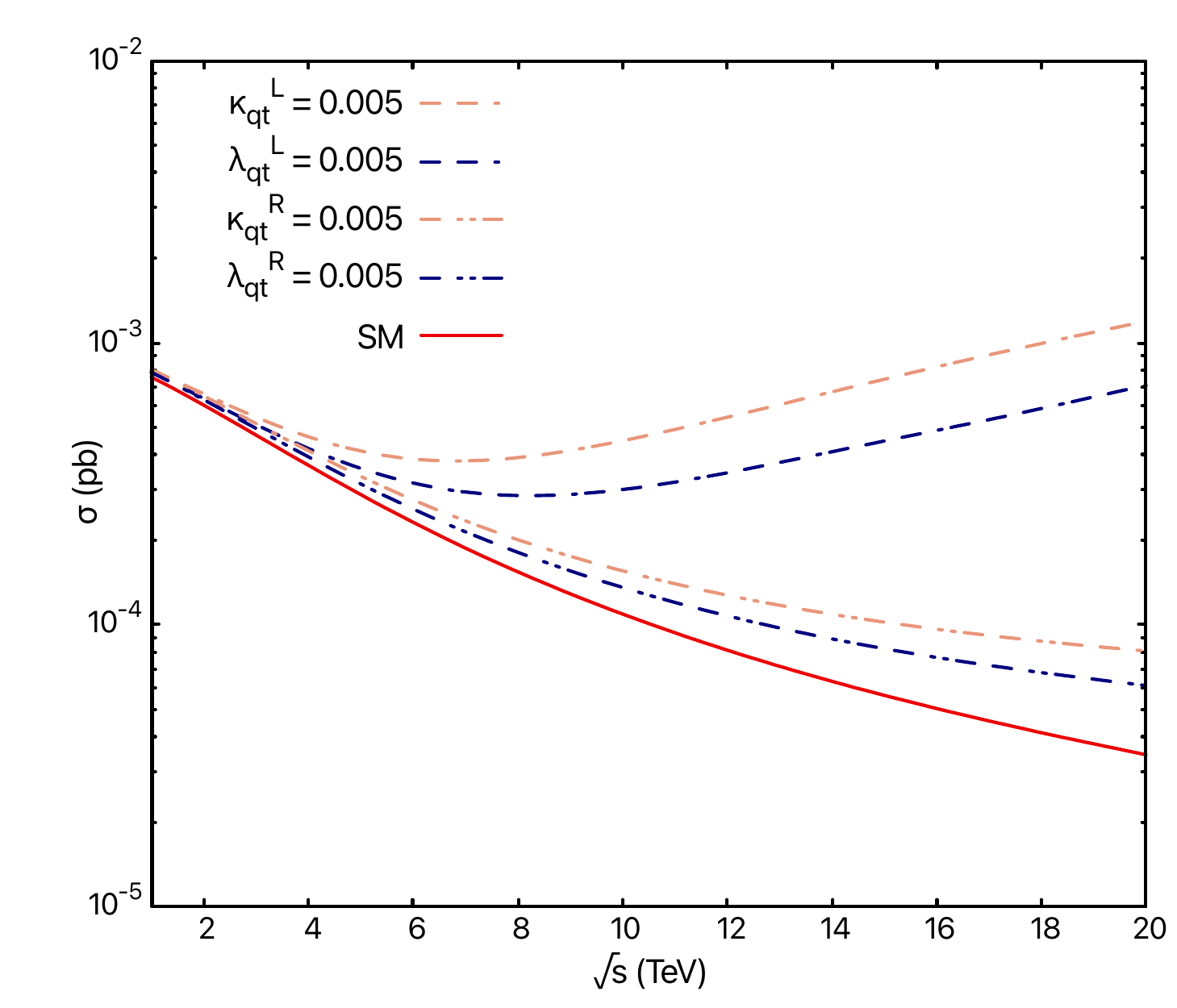}
    \includegraphics[scale=0.32]{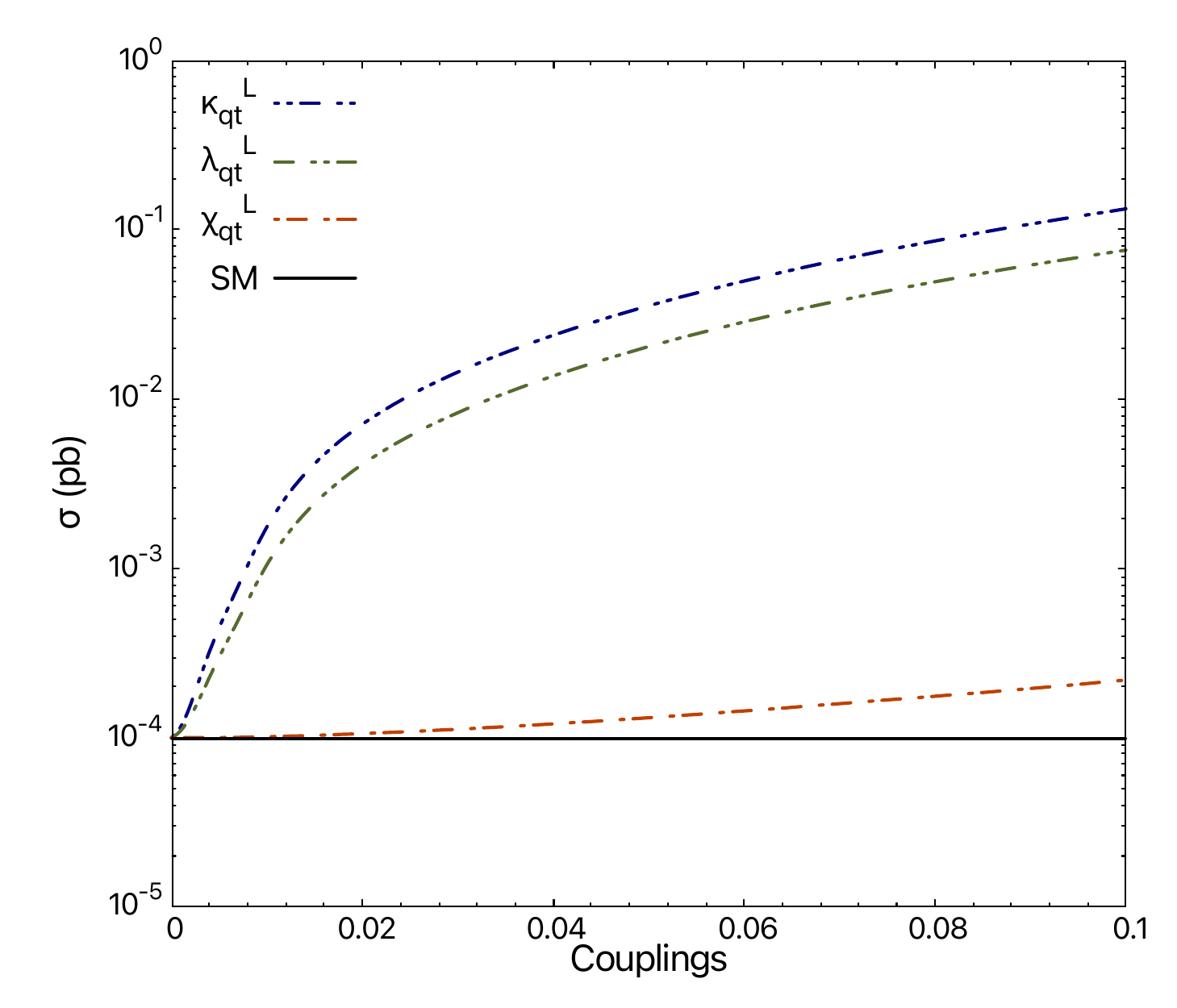}
\caption{Dependence of the total cross section for $\mu^+\mu^- \to \nu_\mu \mu^+ b j$ on the anomalous FCNC couplings defined in Eq.~(1). \textbf{(Left)} Comparison of the left- and right-handed $\kappa_{qt}^{L,R}$ and $\lambda_{qt}^{L,R}$ couplings as a function of $\sqrt{s}$. \textbf{(Right)} Comparison of the three left-handed FCNC interactions, $X_{qt}^L$, $\kappa_{qt}^L$, and $\lambda_{qt}^L$ at  a muon collider with fixed center-of-mass energy of $\sqrt{s}=10~\mathrm{TeV}$. 
}
\label{fig:cross_sections}
\end{figure}
The representative Feynman diagrams contributing to the signal process $\mu^+\mu^- \to \nu_\mu \mu^+ b j$ via anomalous FCNC  $tqZ$ and $tq\gamma$ interactions are shown in Fig.~\ref{fd}. In these diagrams, the top quark is produced through an FCNC vertex involving a neutral gauge boson ($\gamma$ or $Z$) and subsequently decays via the Standard Model charged current channel $t \to W^+ b$, with the $W^+$ boson decaying leptonically. The presence of the FCNC vertex, indicated by the effective interaction in Eq.~(\ref{eq1}), leads to the characteristic final state topology considered in this analysis. In addition to the signal contributions, the same final state can be produced through a variety of Standard Model processes that do not involve FCNC interactions. These background contributions arise from multiple electroweak and single-top production channels, leading to a large number of Feynman diagrams with similar final-state signatures. Due to their complexity and abundance, the corresponding Standard Model diagrams are not shown here; however, they are fully taken into account in the numerical analysis.

To quantitatively evaluate these contributions, the effective FCNC Lagrangian defined in Eq.~(\ref{eq1}) is implemented into the event generator framework using \texttt{FeynRules}~\cite{Alloul:2013bka}, from which a corresponding Universal FeynRules Output (UFO) model~\cite{Degrande:2011ua} is generated. This implementation enables the simulation of the signal process within a Monte Carlo environment, allowing for a consistent incorporation of the anomalous interactions in the event generation. Based on this implementation, the parton-level cross sections for the subprocess $\mu^{+}\mu^{-} \to \nu_{\mu}\,\mu^+\,b\,j$ are computed using \texttt{MadGraph5\_aMC@NLO} (v3.5.4) \cite{Alwall:2014hca} at a muon collider with center-of-mass energy of $\sqrt{s}=10~\mathrm{TeV}$ with the default generator level kinematic cuts. Jets are required to satisfy $p_{T}^{j} > 20~\mathrm{GeV}$, while charged leptons must fulfill $p_{T}^{\ell} > 10~\mathrm{GeV}$. No transverse momentum threshold is imposed on $b$-jets. The pseudorapidity requirements are defined as $|\eta^{j}|<5.0$ for jets and $|\eta^{\ell}|<2.5$ for charged leptons, whereas no upper bound is applied to the pseudorapidity of $b$-jets. Furthermore, an angular separation criterion of $\Delta R_{j\ell} > 0.4$ is imposed to ensure well-isolated objects, while no minimum separation is required between the $b$-jet and the jet or between the $b$-jet and the charged lepton.

To investigate the relative impact of the different FCNC interactions introduced in Eq.~(\ref{eq1}), we consider the most general effective Lagrangian without imposing any  a priori relation between the left- and right-handed couplings. In particular, the tensor couplings are treated independently as $\kappa_{qt}^{L}\neq\kappa_{qt}^{R}$ and $\lambda_{qt}^{L}\neq\lambda_{qt}^{R}$ couplings, while the vector interaction parameterized by $X_{qt}^{L,R}$ is also included in the analysis.

The left panel of Fig.~\ref{fig:cross_sections} shows the dependence of the production cross section for $\mu^{+}\mu^{-}\rightarrow\nu_{\mu}\mu^{+}bj$ on the center-of-mass energy for representative benchmark values of the anomalous couplings. A pronounced difference between the two chiral structures is observed. The contributions induced by the left-handed $\kappa_{qt}^{L}$ and $\lambda_{qt}^{L}$ couplings increase significantly with increasing $\sqrt{s}$, whereas the corresponding right-handed contributions remain strongly suppressed over the entire energy range considered. This behavior originates from the purely left-handed charged-current structure of the Standard Model $tbW$ interaction. In the case of right-handed FCNC vertices, the produced top quark must undergo a chirality flip before coupling to the $W$ boson, leading to a suppression proportional to $m_t^2/s$, which becomes significantly important at high energies. 

Having established the dominance of the left-handed interactions, the right panel of Fig.~\ref{fig:cross_sections} compares the contributions of the left-handed $X_{qt}^{L}$, $\kappa_{qt}^{L}$, and  $\lambda_{qt}^{L}$ couplings. Although the $X_{qt}^{L}$ coupling contributes to the signal process, its effect is found to be substantially smaller than those of the $\kappa_{qt}^{L}$, and  $\lambda_{qt}^{L}$ couplings throughout the range considered. This hierarchy reflects the different Lorentz structures of the effective operators and their corresponding energy dependence. Consequently, the detector-level simulation, multivariate analysis, and sensitivity projections presented in the remainder of this work are restricted to the left-handed $\kappa_{qt}^{L}$ and $\lambda_{qt}^{L}$ couplings, which provide the dominant sensitivity to FCNC top-quark effects at a high-energy muon collider.

\begin{table}[h!]
\centering
\caption{The total cross section ($\sigma$) of the $\mu^{+}\mu^{-} \to \nu_{\mu}\,\mu^+\,b\,j$ process with non-zero couplings ($\kappa^{L}_{qt}=0.0015$ \& $\lambda^{L}_{qt}=0.0030$) and relevant Standard Model backgrounds at $\sqrt{s}=10$ TeV Muon Collider.}\label{cross_section}
\begin{ruledtabular}
\begin{tabular}{l l l c}
 & Processes & Symbol & $\sigma$ [pb] \\
\hline\hline
Signal ($\kappa^{L}_{qt}=0.0015$ \& $\lambda^{L}_{qt}=0.0030$) & $\mu^{+}\mu^{-} \to \nu_{\mu}\mu^+ b j$ & $\nu_{\mu}\mu b j$ ($\kappa^{L}_{qt}(\lambda^{L}_{qt})=0.0015(0.0030)$) & \num{1.86e-4} \\
\hline
& $\mu^{+}\mu^{-} \to$ $\nu_{\mu}\mu^+ b j$ & $\nu_{\mu}\mu b j$ & \num{1.01e-4} \\
& $\mu^{+}\mu^{-} \to$ $\nu_{\mu}\mu^+ b \bar t$ & $\nu_{\mu}\mu b t$ & \num{1.61e-3} \\
& $\mu^{+}\mu^{-} \to$ $t\bar t$ & tt & \(1.79 \times 10^{-3}\) \\
Backgrounds & $\mu^{+}\mu^{-} \to$ $W^+W^-$ & $WW$ & \(5.73 \times 10^{-2}\) \\
& $\mu^{+}\mu^{-} \to$ $W^+W^-\gamma$ & WW$\gamma$ & \(5.15 \times 10^{-3}\) \\
& $\mu^{+}\mu^{-} \to$ $W^+W^-Z$ & WWZ & \(8.94 \times 10^{-3}\) \\
& $\mu^{+}\mu^{-} \to$ $ZZ$ & ZZ & \(3.14 \times 10^{-3}\) \\
& $\mu^{+}\mu^{-} \to$ Z$\gamma$ & Z$\gamma$ & \(4.21 \times 10^{-3}\) \\
\end{tabular}
\end{ruledtabular}
\end{table}

\section{Event Generation and Selection for signal and background processes}\label{section3}
The signal process under consideration proceeds via anomalous FCNC $tqZ$  and $tq\gamma$ interactions, leading to $\mu^{+}\mu^{-} \to \nu_{\mu}\,\mu^+\,b\,j$  production process. The one light-jet, one b-jet, one muon and MET give rise to the characteristic final-state signature. In addition to the signal, the same final state can be generated through irreducible Standard Model processes ($\nu_{\mu}\mu b j$). Furthermore, several reducible backgrounds
same as the signal signature, including diboson production ($WW$, $ZZ$, $Z\gamma$, $WW\gamma$, $WWZ$) and top-related processes such as pair ($tt$) and single top production ($\nu_{\mu}\mu b t$), are considered in the analysis. The total cross sections for the signal and relevant Standard Model background processes, evaluated with the default kinematic cuts of \texttt{MadGraph} mentioned above, are summarized in Table~\ref{cross_section}. While these results provide a first indication of the relative production rates, a more realistic assessment of the signal observability requires a detailed event-level analysis with optimized selection criteria.

To this end, Monte Carlo event samples are generated for all relevant Standard Model background processes and for signal across a two-dimensional $5\times 5$ benchmark grid of non-zero anomalous couplings, $\kappa_{qt}^{L}$ and $\lambda_{qt}^{L}$, spanning five distinct values for each coupling as motivated by the cross-section dynamics shown in Fig.~\ref{fig:cross_sections}. For each process, $2\times10^{6}$ events are generated to ensure sufficient statistical precision in the subsequent analysis. Both signal and background samples are interfaced with \texttt{Pythia}  8.2 \cite{Sjostrand:2014zea} for parton showering and hadronization, followed by a fast detector simulation using \texttt{Delphes} 3.5.6 \cite{deFavereau:2013fsa} software with the dedicated 10 TeV muon collider detector card, namely \texttt{delphes$\_$card$\_$MUSICDet$\_$target.tcl}. Jets are reconstructed by using clustered energy deposits with {\sc FastJet 3.3.2} \cite{Cacciari:2011ma} using anti-kt algorithm \cite{Cacciari:2008gp} where a cone radius is set as $\Delta R$ = 0.4 and $p_T^j>$ 25 GeV.
\begin{table}[h!]
    \caption{Event selection and applied kinematic cuts used in the analysis for the $\mu^{+}\mu^{-} \to \nu_{\mu}\,\mu^+\,b\,j$ process at the $\sqrt{s}=10$ TeV Muon Collider before the multivariate analysis .}
    \begin{ruledtabular}
\begin{tabular}{ll} 
 & Parameters \\ 
 \hline\hline
 Events Pre-selection  & $N^{j} > 0 \ \& \; N^{b} > 0 \ \&  \; N^{\mu} = 1  \ \& \; N^{e} = 0$ \\ 
 & $p_{T}^{\mu} > 20 $ GeV, $p_{T}^{j} > 20 $ GeV, $p_{T}^{b} > 20 $ GeV  \\
 Kinematic Cuts \hspace{5cm} & $|\eta^{\mu}| < 2.5$, $|\eta^{j}| < 2.5$, $|\eta^{b}| < 2.5$, $\slashed{E}_T > 20$~GeV  \\
 & $\Delta R(\mu,b) > 0.4$, $\Delta R(\mu,j) > 0.4$, $\Delta R(j,b) > 0.4$  \\
BDT cuts& $|\cos\theta^{\mu}| < 0.95$, $|\cos\theta^{j}| < 0.95$, $|\cos\theta^{b}| < 0.95$ \\
\end{tabular}
\end{ruledtabular}
\label{table:data}
\end{table}

Starting from these simulated datasets, a sequence of pre-selection requirements and kinematic cuts is applied in order to enhance the signal significance and suppress the Standard Model backgrounds. The details of the event selection strategy are summarized in Table~\ref{table:data}. The analysis begins with a set of pre-selection requirements, including at least one jet and one $b$-tagged jet, exactly one muon, and no electrons in the final state. These criteria are designed to match the characteristic topology of the signal process $\mu^{+}\mu^{-} \to \nu_{\mu}\,\mu^+\,b\,j$, while significantly reducing backgrounds with different lepton multiplicities or flavor compositions.

In particular, purely electroweak diboson processes such as $WW$, $ZZ$, and $Z\gamma$ are partially suppressed at this stage due to the requirement of a $b$-tagged jet, which is not intrinsic to these channels. Similarly, the $WW\gamma$ and $WWZ$ backgrounds are also substantially reduced after the event selection. Owing to their different production topology, these processes have a much lower probability of satisfying the full signal selection requirements, including the reconstruction of a top-quark candidate from a $b$-tagged jet and a light-flavor jet. In contrast, top-related backgrounds such as $t\bar{t}$ and single-top production ($\nu_{\mu}\mu b t$) remain more challenging, since they naturally contain real $b$-jets and leptons, closely resembling the signal topology. Subsequently, a set of kinematic cuts is applied to further enhance the signal-to-background ratio. Transverse momentum requirements on the muon, jet, and $b$-jet suppress soft contributions from multiboson and radiative processes. The pseudorapidity constraints $|\eta|<2.5$ restrict the analysis to the central detector region, improving reconstruction performance and reducing forward background contamination. The requirement of missing transverse energy, $\slashed{E}_T > 20$~GeV, plays a crucial role in rejecting backgrounds without genuine neutrinos, particularly $Z\gamma$ and $ZZ$ processes where missing energy arises primarily from detector effects. Furthermore, angular separation criteria of $\Delta R>0.4$ are imposed between all reconstructed final-state objects, together with the detector acceptance requirement $|\cos\theta|<0.95$. These selections reduce the impact of collinear configurations and object overlap, while ensuring reliable lepton and jet reconstruction within the fiducial detector region. After the full selection, the dominant residual backgrounds are expected to arise from $t\bar{t}$ and single-top processes, due to their similar final-state structure. Although the cut-based selection significantly improves the signal-to-background ratio, it is not sufficient for optimal separation. This motivates the use of multivariate analysis techniques in the subsequent stage to fully exploit the kinematic differences between signal and background events.
\section{the details of the multivariate analysis with boosted decision tree}\label{section4}
To enhance the sensitivity to the $\nu_{\mu} \mu^+ b j$ signal process, particularly for small coupling values, and to effectively suppress the Standard Model (SM) backgrounds remaining after the kinematic cuts in Table~\ref{table:data}, a Boosted Decision Tree (BDT) classifier is implemented for each signal scenario. The BDT is trained using a combination of object-level kinematic observables, such as transverse momenta ($p_T$), pseudorapidities ($\eta$), and angular separations, alongside reconstructed system-level variables including the invariant mass $m_{bj}$ and the total transverse momentum $p_T^{bj}$. Using the Toolkit for Multivariate Data Analysis (TMVA) default hyperparameter configuration, an initial training is performed to rank the discriminating power of each input observable.  To ensure a robust and universal feature set, the 24 most influential variables are identified for each coupling scenario, and only those common to all scenarios are retained for the final BDT training and hyperparameter optimization.

The relative importance of input variables in a BDT is determined based on their contribution to the classifier's decision. 
In the default TMVA BDT implementation, this importance is calculated using Gini impurity reduction. 
The Gini impurity, which measures the “mixedness” of a node in a decision tree, is defined as
\begin{equation}
G = 1 - \sum_{k=1}^{K} p_k^2
\end{equation}
where $G$ is the Gini impurity of the node, $K$ is the number of classes in the node, and $p_k$ is the fraction of samples in the node belonging to class $k$. 
A node is completely pure ($G=0$) if all samples belong to a single class, while the impurity is maximal when the samples are equally distributed among all classes. To quantify the discriminating power of the selected observables, Table~\ref{tab:table2} shows the variable importance obtained from a representative benchmark point with $\kappa^{L}_{qt}=0.0015$ and $\lambda^{L}_{qt}=0.0030$ at $\sqrt{s}=10$~TeV. This benchmark is chosen for illustration purpose only. We have verified that the relative ordering and discriminating performance of the input variables are stable throughout the anomalous coupling parameter space explored in this work, indicating that the conclusions drawn from Table~\ref{tab:table2} are generally applicable to the full analysis. 

\begin{table}[htb!]
\caption{The input variable list of the BDT: the kinematic and reconstructed variables from the leading muon ($\mu$), b-tagged jet (b) and light jet (j) and Missing transverse energy. The relative importance of the input variables for the $\mu^+\mu^- \to \nu_\mu \mu^+ b j$ process at a benchmark coupling value of Signal ($\kappa^{L}_{qt}=0.0015$ \& $\lambda^{L}_{qt}=0.0030$) for $\sqrt{s} = 10~\text{TeV}$ at the Muon Collider.}
\begin{ruledtabular}
\begin{tabular}{llc}
Variable & Definition& Relative Importance\\
 &  & [\%] \\
 \hline
$p_{T}^{j}$ & Transverse momentum of the leading light jet & 7.10 \\
$M_{bj}$ & Invariant mass of reconstructed $bj$ system & 6.47 \\
$p_{T}^{\mu}$ & Transverse momentum of the leading muon & 6.22 \\
$\cos\theta^{\mu b}$ & Cosine of the polar angle between leading muon and b-tagged jet & 5.81 \\
$p_{T}^{bj}$ & Transverse momentum between leading b-tagged jet and light jet & 5.75 \\
$N^{\text{b}}$ & Number of b-tagged jets in the event & 5.45 \\
$p_{T}^{b}$ & Transverse momentum of the leading b-tagged jet & 5.39 \\
$\Delta R({\mu, b})$ & Distance between leading muon and leading b-tagged jet in $\eta$-$\phi$ plane & 5.16 \\
$N^{\text{j}}$ & Number of light jets in the event & 5.05 \\
$M_T$ & Transverse mass of the $\nu\mu bj$ system & 4.60 \\
$\cos\theta^{\mu j}$ & Cosine of the polar angle between leading muon and light jet & 4.22 \\
$\slashed{E}_T$ & Missing transverse energy & 4.20 \\
$\cos\theta^{bj}$ & Cosine of the polar angle between leading b-tagged jet and light jet & 4.14 \\
$\Delta R({b,j})$ & Distance between leading b-tagged jet and light jet in $\eta$-$\phi$ plane & 3.75 \\
$\Delta R({\mu, j})$ & Distance between leading muon and light jet in $\eta$-$\phi$ plane & 3.54 \\
$\cos\theta^{j}$ & Cosine of the polar angle of the leading light jet & 2.80 \\
$\eta^{b}$ & Pseudorapidity of the leading b-tagged jet & 2.69 \\
$\cos\theta^{\mu}$ & Cosine of the polar angle of the leading muon & 2.69 \\
$\eta^{j}$ & Pseudorapidity of the leading light jet & 2.66 \\
$\phi^{b}$ & Azimuthal angle of the leading b-tagged jet & 2.63 \\
$\phi^{\mu}$ & Azimuthal angle of the leading muon & 2.58 \\
$\cos\theta^{b}$ & Cosine of the polar angle of the leading b-tagged jet & 2.45 \\
$\eta^{\mu}$ & Pseudorapidity of the leading muon & 2.36 \\
$\phi^{j}$ & Azimuthal angle of the leading light jet & 2.32 \\
\end{tabular}
\end{ruledtabular}
\label{tab:table2}
\end{table}

The ranking demonstrates that the transverse momenta of final-state particles and their angular correlations provide the most significant discrimination between the anomalous signal contributions and the SM background. Particularly, the invariant mass of the 
reconstructed light jet and the leading b-tagged jet, $M_{bj}$ system, provides the strongest 
discrimination between signal and backgrounds. The transverse momenta of the leading light jet, b-tagged jet and muon ($p_T^{j}$, $p_T^{b}$, $p_T^{\mu}$), together with the missing transverse mass of $\nu\mu bj$ system, also provide substantial discrimination power. Moreover, angular observables such as $\Delta R(b,j)$, $\Delta R(\mu,b)$, and $\Delta R(\mu,j)$, along with $\cos\theta$ variables between different particle combinations, provide additional sensitivity by capturing the characteristic spatial configurations of the final-state particles in signal events.

\begin{figure}[httt!]
    \centering
    \includegraphics[scale= 0.4 ]{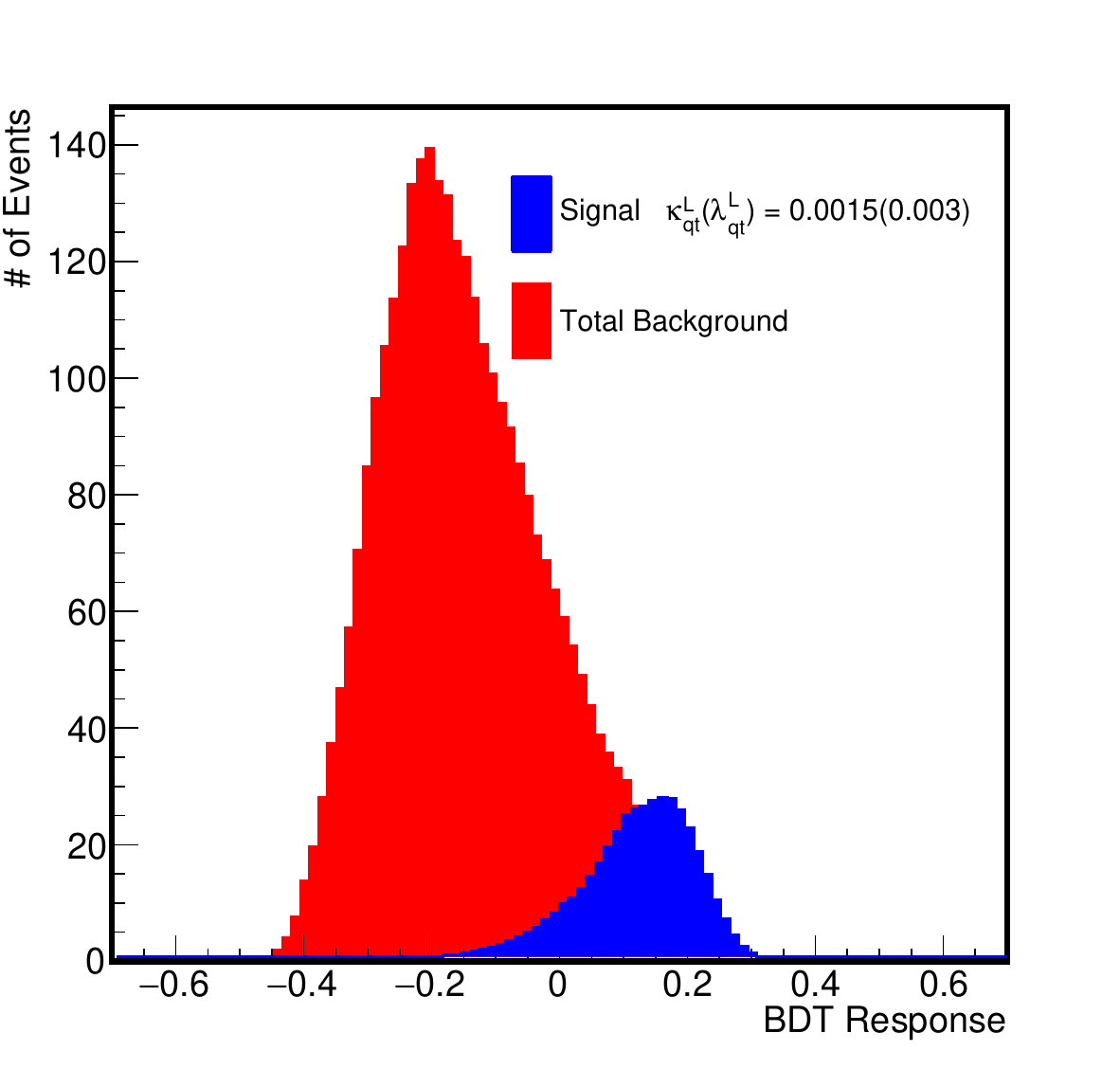}
    \includegraphics[scale= 0.4 ]{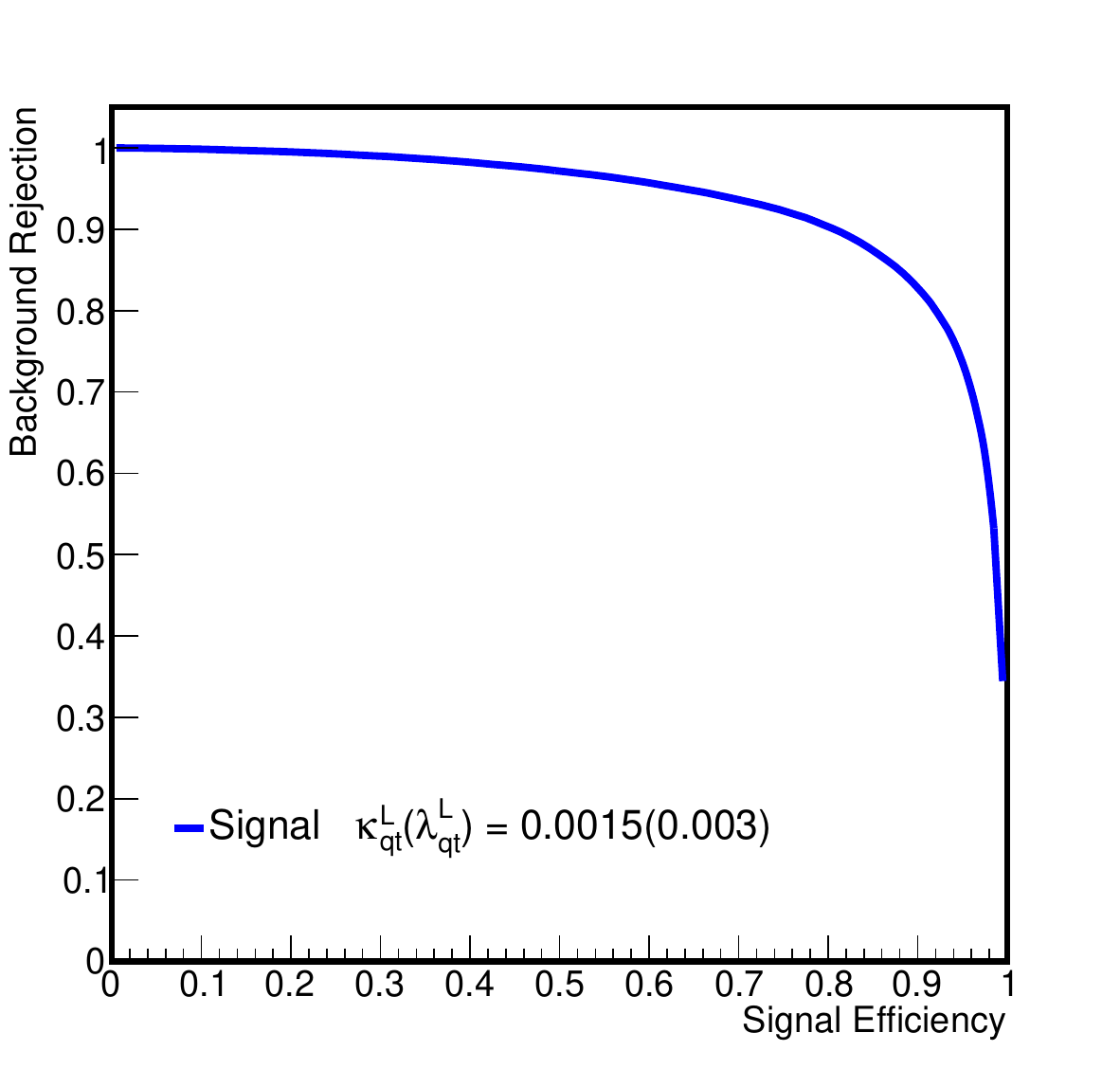}
    \caption{The left panel show the BDT output for the signal sample ($\kappa^{L}_{qt}=0.0015$ \& $\lambda^{L}_{qt}=0.0030$) together with the total background. The right panel present the corresponding ROC curves obtained from the same training.}
    \label{fig_roc}
\end{figure}

The training is performed using adaptive boosting (AdaBoost) with 450 decision trees, a maximum tree depth of 3, a minimum of 2.5 events per terminal leaf, 20 cut optimization steps, and a learning rate of 0.5. The full event sample is divided into statistically independent subsets, with 50\% of the data used for training and the remaining 50\% reserved for testing. The performance of the boosted decision tree (BDT) classifier is illustrated in Fig.~\ref{fig_roc} for representative benchmark scenarios with ($\kappa^{L}_{qt}=0.0015$ \& $\lambda^{L}_{qt}=0.0030$) (on the left). The left panel show the BDT response distributions for the signal and the total background. A clear separation between the two can be observed, with signal events predominantly populating the positive BDT response region, while background events are concentrated at lower (negative) values. This behavior indicates that the BDT successfully captures the underlying kinematic differences between signal and background processes. The corresponding receiver operating characteristic (ROC) curves are presented in the right panel of Fig.~\ref{fig_roc}. These curves demonstrate the trade-off between signal efficiency and background rejection achieved by the classifier. A high level of background rejection is maintained over a wide range of signal efficiencies, indicating a strong discriminating power of the BDT. In particular, the relatively slow degradation of background rejection with increasing signal efficiency reflects the robustness of the multivariate approach.  Overall, these results confirm that the BDT-based multivariate analysis provides an efficient tool for enhancing the signal sensitivity and suppressing the dominant Standard Model backgrounds.
\begin{table}[h!]
\centering  
\caption{Normalized number of events after applied cuts  given in Table~\ref{table:data} for $\mu^{+}\mu^{-} \to \nu_{\mu}\,\mu^+\,b\,j$ signal process with non-zero couplings ($\kappa^{L}_{qt}=0.0015$ \& $\lambda^{L}_{qt}=0.0030$) and  relevant Standard Model backgrounds (SM) at the $\sqrt{s}=10$ TeV Muon Collider.}
\begin{ruledtabular}
    \begin{tabular}{l l c c c} 
 & Processes & Pre-Selection & Kinematic  & BDT \\ [0.5ex] 
 \hline\hline
Signal~($\kappa^{L}_{qt}=0.0015$ \& $\lambda^{L}_{qt}=0.0030$) & $\mu \nu_{\mu} bj$  & $6.560 \times 10^{2}$ & $5.862 \times 10^{2}$ & $3.985 \times 10^{2}$ \\ \hline
  & $\mu \nu_{\mu} bj$ & $1.374 \times 10^{2}$ & $1.168 \times 10^{2}$ & $2.340 \times 10^{1}$ \\
  & $\mu \nu_{\mu} bt$ & $6.511 \times 10^{3}$ & $5.516 \times 10^{3}$ & $3.819 \times 10^{2}$ \\
 Backgrounds & tt & $1.663 \times 10^{3}$ & $7.407 \times 10^{2}$ & $1.146 \times 10^{2}$ \\
  & WWZ & $1.048 \times 10^{3}$ & $6.932 \times 10^{2}$ & $1.376 \times 10^{2}$ \\ 
  & WW$\gamma$ & $5.759 \times 10^{1}$ & $2.866 \times 10^{1}$ & $5.541 \times 10^{0}$ \\ 
  & WW & $6.475 \times 10^{1}$ & $9.357 \times 10^{0}$ & $1.528 \times 10^{0}$
\end{tabular}
\end{ruledtabular}
\label{noevent_kappa}
\end{table}

The impact of the event selection strategy on both the signal and the background processes is summarized in Table~\ref{noevent_kappa}, where the normalized number of events is presented after each stage of the analysis, including pre-selection, kinematic cuts, and the final BDT-based selection. At the pre-selection level, the requirements on jet multiplicity, $b$-tagging, and lepton content significantly reduce backgrounds with incompatible final-state topologies, while retaining a substantial fraction of the signal events. In particular, electroweak processes such as $WW$ and $WW\gamma$ are already strongly suppressed due to the presence of a $b$-tagged jet requirement, which is not intrinsic to these channels. The application of kinematic cuts further enhances the signal-to-background ratio. Requirements on transverse momentum, pseudorapidity, missing transverse energy, and angular separation effectively suppress soft and forward background contributions. As a result, a noticeable reduction is observed in multiboson backgrounds such as $WWZ$ and $WW\gamma$, while top-related backgrounds, especially $t\bar{t}$ and $\nu_\mu \mu b t$, remain comparatively more resilient due to their similar final-state structure to the signal. To determine the optimal threshold for the reconstructed BDT response distributions, we select the specific cut-off point that maximizes statistical significance. 
 
The final selection based on the BDT classifier leads to a substantial improvement in background rejection while preserving a significant portion of the signal. In particular, the dominant backgrounds are reduced by more than an order of magnitude in some cases, whereas the signal efficiency remains relatively high. This demonstrates the strong discriminating power of the multivariate approach, which efficiently exploits the kinematic differences between signal and background events. Overall, the results presented in Table~IV highlight the progressive enhancement of the signal significance through successive selection steps, with the BDT-based analysis providing the most effective separation between signal and background processes. 
\begin{figure}[httt!]
    \centering
    \includegraphics[scale = 0.4]{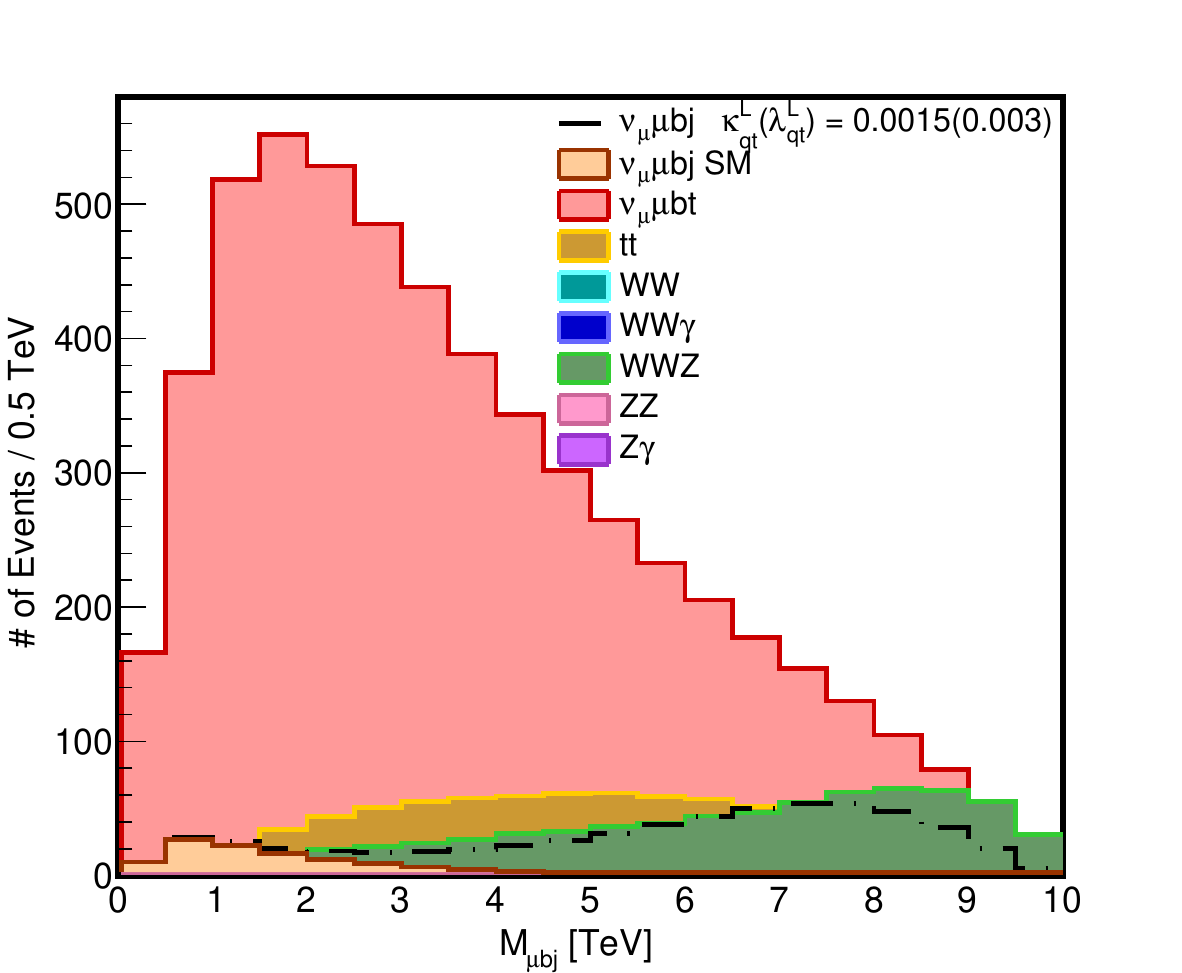}
     \centering
    \includegraphics[scale= 0.4]{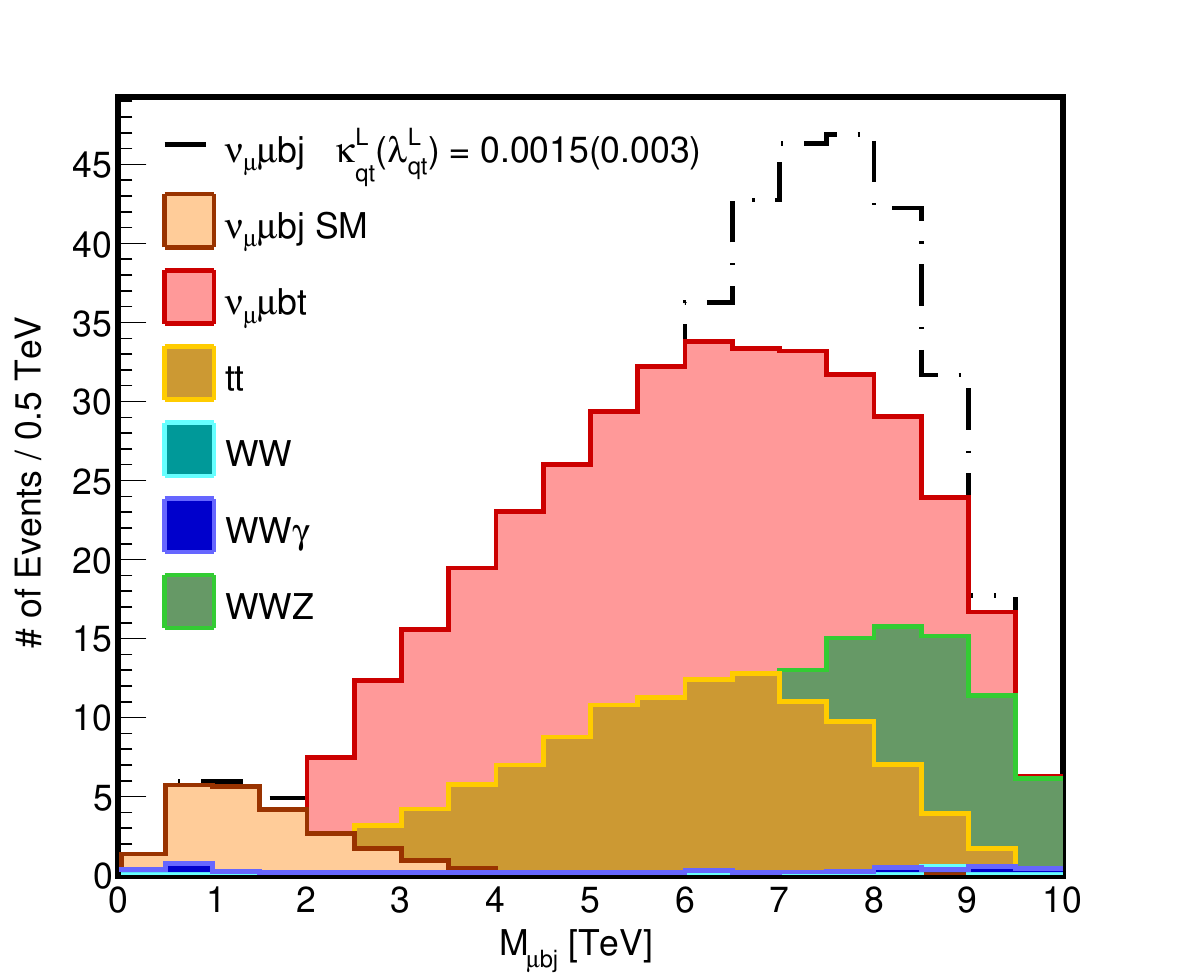}
    \caption{Normalized distribution of the reconstructed invariant mass $M_{\mu bj}$ for the signal process $\nu_{\mu}\mu bj$ with anomalous coupling ($\kappa^{L}_{qt}=0.0015$ \& $\lambda^{L}_{qt}=0.0030$), compared with all relevant Standard Model background processes. The left panel shows the histogram obtained after applying the selection cuts given in Table~II, while the right panel shows the distribution after including the BDT score.}
    \label{minv_mass}
\end{figure}

The normalized invariant mass distribution of the $\mu b j$ system for the signal process $\mu^+\mu^- \to \nu_\mu \mu^+ b j$ with ($\kappa^{L}_{qt}=0.0015$ \& $\lambda^{L}_{qt}=0.0030$), together with the relevant Standard Model backgrounds, is presented in Fig.~\ref{minv_mass}. The left panel shows the distributions after the application of the baseline kinematic selection cuts. At this stage, the background contributions, particularly from $\nu_{\mu}\mu bt$ and $tt$ processes, dominate the spectrum, rendering the signal nearly indistinguishable. The right panel illustrates the same distribution after the application of the BDT-based selection. A substantial suppression of the Standard Model backgrounds is observed, leading to a clear enhancement of the signal contribution. In particular, background processes such as $ZZ$ and $Z\gamma$, which are present at the pre-selection level, are completely eliminated after the multivariate selection due to their negligible event yields. As a result, the signal becomes prominent, especially in the high invariant mass region between $2$ and $10$~TeV. These results highlight that the $M_{\mu b j}$ observable provides strong discriminating power between signal and background processes, with the signal preferentially populating the high-mass region. Such kinematic differences can be systematically exploited to improve the separation between signal and background, either through optimized cut-based selections or, more effectively, via multivariate techniques.

In this context, the invariant mass distribution, along with other kinematic observables, constitutes a powerful input for the statistical interpretation of the analysis. The enhanced signal-to-background ratio achieved after the BDT selection directly translates into an improved sensitivity, motivating a quantitative evaluation of the expected statistical significance, which is discussed in the following section.

\section{Limits on Anomalous $tqZ$ and $tq\gamma$ Couplings}\label{section5}

The projected sensitivities are obtained through separate two-parameter scans of the anomalous FCNC couplings. In the first scan, the real components of the left-handed couplings, $\mathrm{Re}(\kappa_{qt}^{L})$ and $\mathrm{Re}(\lambda_{qt}^{L})$, are varied simultaneously while the imaginary components are fixed to zero. In the second scan, the imaginary components, $\mathrm{Im}(\kappa_{qt}^{L})$ and $\mathrm{Im}(\lambda_{qt}^{L})$, are varied simultaneously with the real components set to zero. The corresponding discovery and exclusion sensitivities are extracted in the two-dimensional parameter planes, allowing the combined effects of the anomalous $tqZ$ and $tq\gamma$ interactions to be explored in each case. The resulting sensitivity contours are presented in the $\mathrm{Re}(\kappa_{qt}^{L})$--$\mathrm{Re}(\lambda_{qt}^{L})$ and $\mathrm{Im}(\kappa_{qt}^{L})$--$\mathrm{Im}(\lambda_{qt}^{L})$ planes for a muon collider operating at $\sqrt{s}=10~\mathrm{TeV}$ with an integrated luminosity of $10~\mathrm{ab}^{-1}$.

The statistical significance for discovery ($SS_{\rm disc}$) and exclusion ($SS_{\rm excl}$) is evaluated using a profile-likelihood--based approach that incorporates the effect of systematic uncertainties. The median expected significances are defined as \cite{Cowan:2010js,Kumar:2015tna}

\begin{equation}
SS_{\rm disc} =
\sqrt{ 2 \left[ (S+B)\ln\!\left(\frac{(S+B)(1+\delta^2 B)}{B+\delta^2 B(S+B)}\right)
- \frac{1}{\delta^2}\ln\!\left(1+\delta^2\frac{S}{1+\delta^2 B}\right) \right] },
\label{eq:ssdisc}
\end{equation}

\begin{equation}
SS_{\rm excl} =
\sqrt{ 2 \left[ S - B \ln\!\left(\frac{B+S+x}{2B}\right)
- \frac{1}{\delta^2}\ln\!\left(\frac{B-S+x}{2B}\right) \right]
- (B+S-x)\!\left(1+\frac{1}{\delta^2 B}\right)},
\label{eq:ssexcl}
\end{equation}
with
\begin{equation}
x = \sqrt{ (S+B)^2 - \frac{4 \delta^2 S B^2}{1+\delta^2 B} } .
\end{equation}
In the limit of negligible systematic uncertainty ($\delta \to 0$), these expressions reduce to the well-known forms
\begin{equation}
SS_{\rm disc} = \sqrt{ 2 \left[ (S+B)\ln(1+S/B) - S \right] },
\end{equation}
\begin{equation}
SS_{\rm excl} = \sqrt{ 2 \left[ S - B \ln(1+S/B) \right] } .
\end{equation}
where $S$ and $B$ denote the expected numbers of signal and total Standard Model background events after the optimized BDT selection. These event yields are obtained by integrating the invariant-mass distribution of the $\mu bj$ system over the interval
$2 < M_{\mu bj} < 10~\mathrm{TeV}$, which defines the signal region used throughout the statistical analysis.
\begin{figure}[httt!]
    \centering
    \includegraphics[scale=0.45]{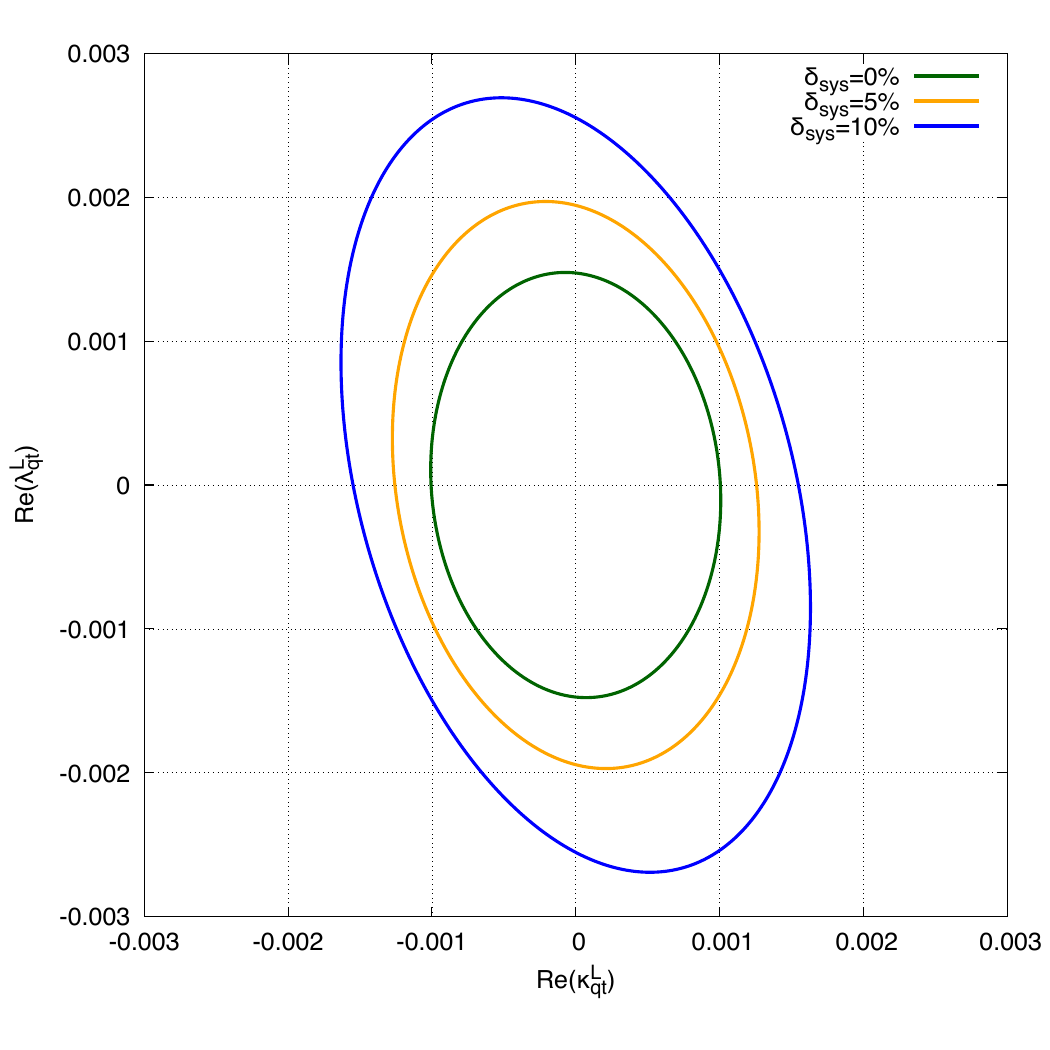}
    \includegraphics[scale=0.45]{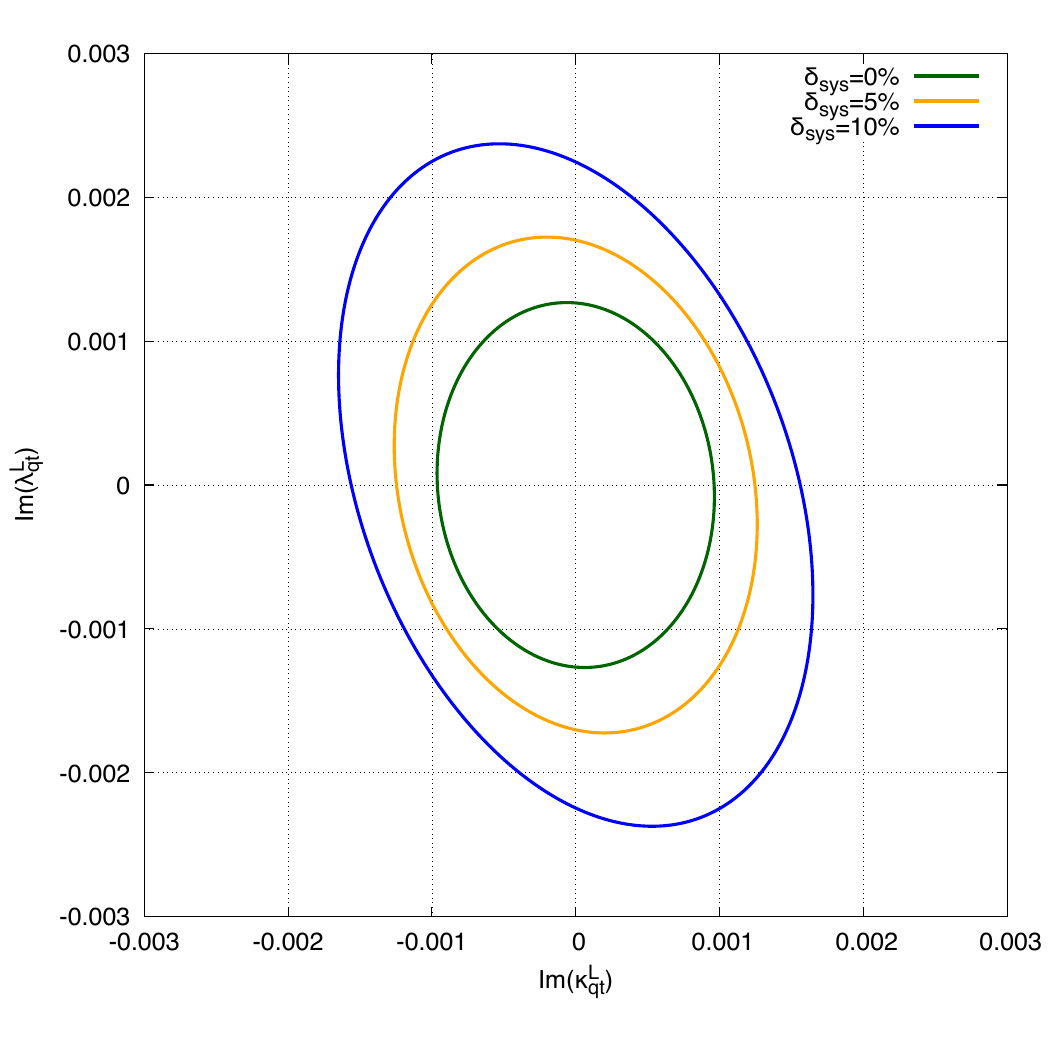}

    \vspace{0.5cm}

    \includegraphics[scale=0.45]{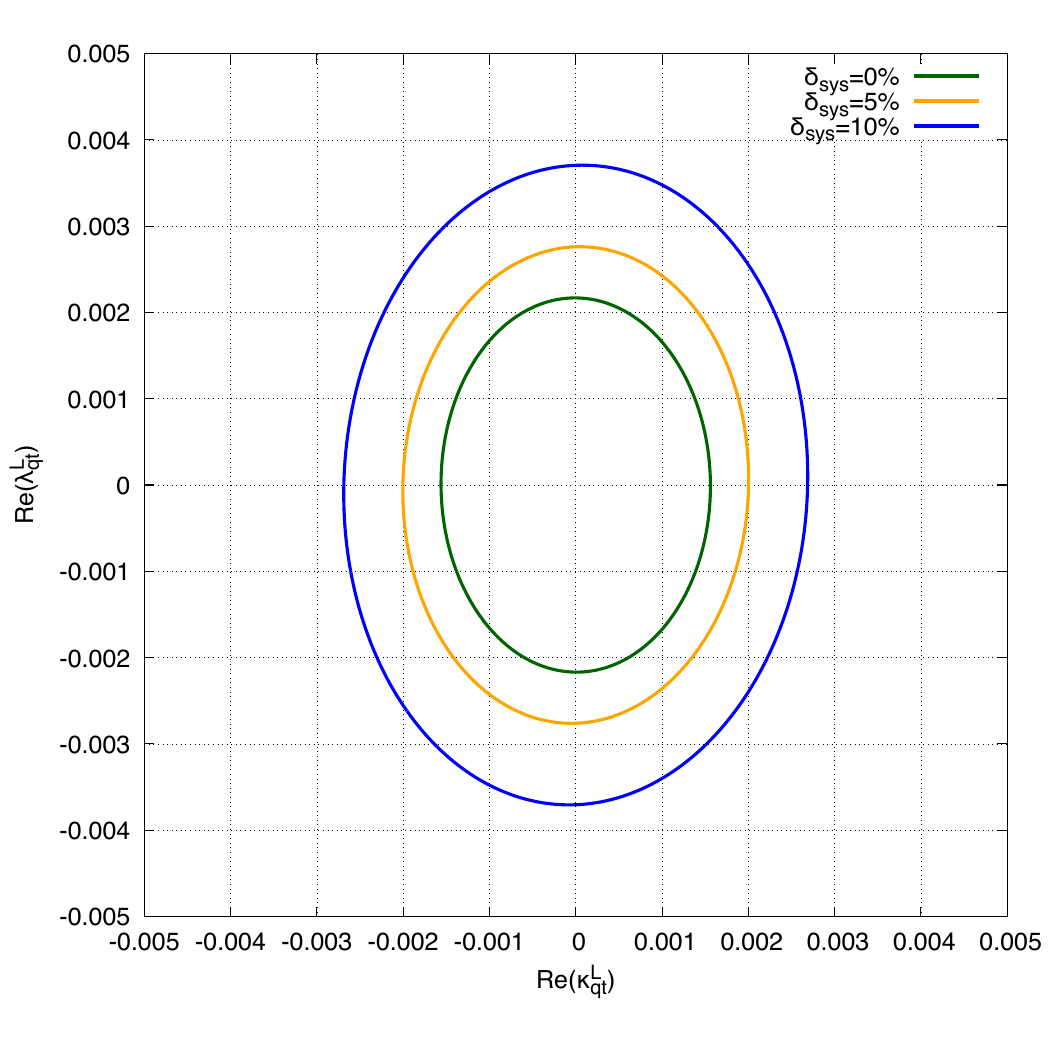}
    \includegraphics[scale=0.45]{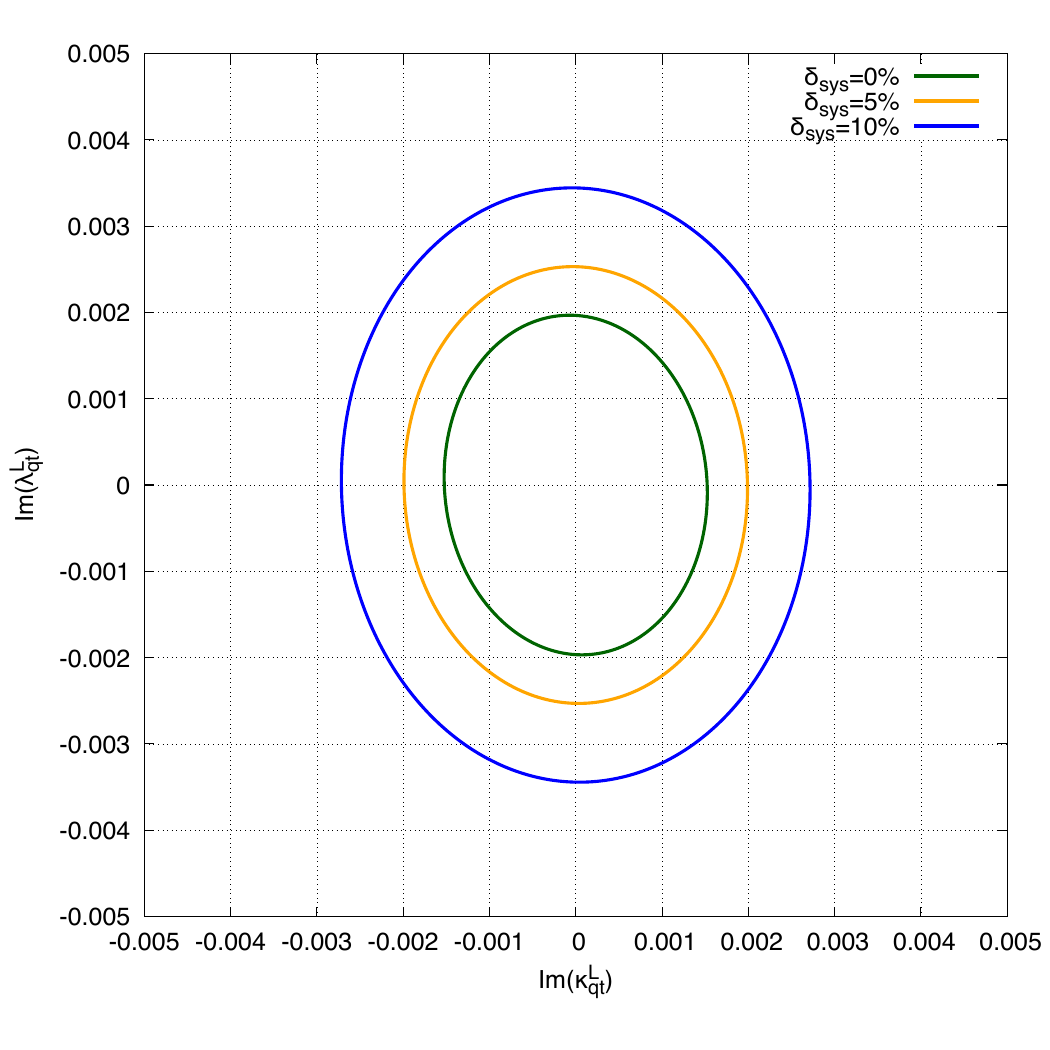}
    \caption{Projected sensitivity contours in the anomalous coupling planes obtained from the statistical significance analysis for the process $\mu^{+}\mu^{-}\rightarrow\nu_{\mu}\mu^{+}bj$ at a $\sqrt{s}=10$ TeV muon collider with an integrated luminosity of $10~\mathrm{ab}^{-1}$. The upper panels show the $95\%$ C.L. exclusion contours derived from $SS_{\rm excl}$, while the lower panels present the corresponding $5\sigma$ discovery contours obtained from $SS_{\rm disc}$. The left (right) column corresponds to the real (imaginary) components of the left-handed anomalous couplings, namely $\mathrm{Re}(\kappa_{tq}^{L})$--$\mathrm{Re}(\lambda_{tq}^{L})$ and $\mathrm{Im}(\kappa_{tq}^{L})$--$\mathrm{Im}(\lambda_{tq}^{L})$, respectively. In each panel, the green, orange, and blue contours represent the systematic uncertainty scenarios of $\delta_{\rm sys}=0\%$, $5\%$, and $10\%$, respectively.}
    \label{fig:ss_combined}
\end{figure}

In order to provide a realistic assessment of the collider sensitivity, the impact of systematic uncertainties is explicitly taken into account. The dominant sources include uncertainties associated with the background normalization, integrated luminosity, lepton identification efficiency, $b$-tagging performance, and jet reconstruction. To quantify their effect, three benchmark scenarios are considered, corresponding to relative systematic uncertainties of $\delta_{\rm sys}=0\%,\,5\%,$ and $10\%$.

The combined sensitivities to the anomalous FCNC couplings are presented in
Fig.~\ref{fig:ss_combined}, where the projected exclusion and discovery reaches are shown in the two-dimensional parameter planes obtained from the
simultaneous variation of the left-handed couplings $\kappa_{tq}^{L}$ and $\lambda_{tq}^{L}$. The upper panels display the $95\%$ C.L. exclusion contours derived from the profile-likelihood estimator $SS_{\rm excl}$, while the lower panels show the corresponding $5\sigma$ discovery contours obtained from $SS_{\rm disc}$. The left and right columns illustrate the
sensitivities in the $\mathrm{Re}(\kappa_{tq}^{L})$--$\mathrm{Re}(\lambda_{tq}^{L})$ and $\mathrm{Im}(\kappa_{tq}^{L})$--$\mathrm{Im}(\lambda_{tq}^{L})$ parameter planes, respectively.

The resulting contours are approximately elliptical and are centered around the origin, as expected in the absence of interference with the Standard
Model amplitude. The elongation and orientation of the ellipses reflect the relative contributions of the anomalous $tqZ$ and $tq\gamma$ interactions and their mutual interference. In particular, the small rotation of the principal axes with respect to the coordinate axes indicates a nonzero correlation between the two anomalous couplings. The full contours are shown
over the entire parameter plane, allowing both their principal-axis orientation and their extremal points to be directly identified. The same physical scale is used for the horizontal and vertical axes in each panel,
which facilitates a direct comparison of the relative sensitivities to the two couplings.

The influence of systematic uncertainties is clearly visible in all parameter planes. As $\delta_{\rm sys}$ increases from $0\%$ to $5\%$ and $10\%$, the contours expand systematically toward larger coupling values, indicating a gradual degradation of the achievable sensitivity. Nevertheless, even for the conservative assumption of a $10\%$ systematic uncertainty, the projected sensitivities remain at the level of a few $10^{-3}$ for the effective couplings. The similarity of the contour sizes and shapes between the real and imaginary coupling planes further indicates that the overall sensitivity is  primarily determined by the magnitude of the anomalous interactions, with only a moderate dependence on the complex phase.

The numerical sensitivities extracted from the two-dimensional contour analyses are summarized in Table~\ref{tab:ss_limits}. The values listed in the table correspond to the individual limits, obtained from the  intersections of the exclusion ($95\%$ C.L.) and discovery ($5\sigma$) contours with the coordinate axes in the $\mathrm{Re}(\kappa_{qt}^{L})$--$\mathrm{Re}(\lambda_{qt}^{L})$ and $\mathrm{Im}(\kappa_{qt}^{L})$--$\mathrm{Im}(\lambda_{qt}^{L})$ parameter planes, with the other anomalous coupling fixed to zero. Even though the contours extends symmetrically resulting both positive and negative coupling values, only absolute values of the individual limits are reported in the Table~\ref{tab:ss_limits}. Consequently, the values reported outside the square brackets represent the limits on the real components of the anomalous couplings, whereas those given in square brackets correspond to the imaginary components. In the absence of systematic uncertainties, the left-handed tensor couplings can be probed down to approximately  $|\mathrm{Re}(\kappa_{qt}^{L})|=1.56\times10^{-3}$ and $|\mathrm{Im}(\kappa_{qt}^{L})|=1.52\times10^{-3}$ at the $5\sigma$ discovery level. For the $tq\gamma$ interaction, the corresponding sensitivities are $|\mathrm{Re}(\lambda_{qt}^{L})|=2.17\times10^{-3}$ and $|\mathrm{Im}(\lambda_{qt}^{L})|=1.97\times10^{-3}$. As expected, increasing the assumed systematic uncertainty gradually shifts the sensitivity toward larger coupling values.

We have also examined the corresponding marginal limits obtained from the extremal points of the full two-dimensional contours, where both anomalous couplings are allowed to vary simultaneously. The marginal limits are found to be very close to the individual limits, with relative differences of up to approximately $5.6\%$ for the $95\%$ C.L. contours and below $0.11\%$ for the $5\sigma$ discovery contours. This indicates that the $\kappa$--$\lambda$ correlations and the associated $Z/\gamma$ interference effects have only a small impact on the extracted one-dimensional coupling limits. The small difference between the individual and marginal limits is more pronounced for the $95\%$ C.L. contours at larger systematic  uncertainties, while it remains very small for the $5\sigma$ discovery contours.
\begin{table}[httt!]
\centering
\caption{Sensitivity limits on the left-handed anomalous FCNC
couplings extracted from the two-dimensional contour analyses shown in
Fig.~\ref{fig:ss_combined}. The quoted values correspond to the absolute
values of the individual limits, obtained from the intersections of the
contour boundaries with the coordinate axes, with the other anomalous
coupling fixed to zero. Numbers outside (inside) square brackets denote the
absolute values of the limits on the real (imaginary) components of the
anomalous couplings.}
\label{tab:ss_limits}
\begin{ruledtabular}
    \begin{tabular}{c c c c }
 Coefficient & $\delta_{\rm sys}$ & $5\sigma$ & 95\% C.L \\ 
 \hline \hline
\multirow{3}{*}{$|\mathrm{Re}(\kappa_{qt}^L)|,[|\mathrm{Im}(\kappa_{qt}^L)|]$}
 & $0\%$   & $1.56\times10^{-3},[1.52\times10^{-3}]$
& $1.01\times10^{-3},[9.62\times10^{-4}]$  \\
 & $5\%$    & $2.00\times10^{-3},[1.99\times10^{-3}]$
& $1.26\times10^{-3},[1.25\times10^{-3}]$ \\
 & $10\%$  & $2.69\times10^{-3},[2.72\times10^{-3}]$
& $1.55\times10^{-3},[1.56\times10^{-3}]$  \\ 
 \hline
\multirow{3}{*}{$|\mathrm{Re}(\lambda_{qt}^L)|,[|\mathrm{Im}(\lambda_{qt}^L)|]$}
 & $0\%$    & $2.17\times10^{-3},[1.97\times10^{-3}]$
& $1.47\times10^{-3},[1.27\times10^{-3}]$\\
 & $5\%$     & $2.76\times10^{-3},[2.53\times10^{-3}]$
& $1.94\times10^{-3},[1.70\times10^{-3}]$ \\
 & $10\%$   & $3.70\times10^{-3},[3.44\times10^{-3}]$
& $2.55\times10^{-3},[2.25\times10^{-3}]$\\ 
\end{tabular}
\end{ruledtabular}
\end{table}

The projected limits on the anomalous couplings can be translated into the corresponding branching ratios of the rare top-quark decays, providing a direct connection between the effective-field-theory parameters and experimentally observable quantities. Since current experimental searches conventionally report their results in terms of branching fractions, this translation enables a straightforward comparison between the projected sensitivities obtained in this work and the existing limits from the ATLAS and CMS Collaborations.

The branching ratios are evaluated using the effective FCNC Lagrangian introduced in Eq.~(2). In the most general case, both left- and right-handed anomalous couplings contribute to the rare decay modes $t\to qZ$ and $t\to q\gamma$ ($q=u,c$). The total top-quark decay width is therefore written as
\begin{equation}
\Gamma_t = \Gamma(t\to W^+b) + \Gamma(t\to W^+s) + \Gamma(t\to W^+d)
+ \Gamma(t\to qZ) + \Gamma(t\to q\gamma),
\end{equation}
where the dominant contribution arises from the Standard Model charged-current decay $t \to W^+ b$.

At leading order, the dominant Standard Model contribution is
\begin{equation}
\Gamma(t\to W^+b) =
\frac{\alpha |V_{tb}|^2}{16 s_w^2}\frac{m_t^3}{m_W^2}
\left(1 - \frac{3 m_W^4}{m_t^4} + \frac{2 m_W^6}{m_t^6}\right).
\end{equation}

On the other hand, the FCNC partial decay widths induced by the effective operators in Eq.~(2) are expressed as
\begin{equation}
\Gamma(t\to q\gamma) = \frac{\alpha}{4}\left((\lambda_{qt}^L)^2 + (\lambda_{qt}^R)^2\right)m_t ,
\end{equation}
\begin{equation}
\Gamma(t\to qZ) =
\frac{\alpha}{32 s_w^2 c_w^2 m_Z^2}
\left((\kappa^L_{qt})^2 + (\kappa_{qt}^R)^2\right)
m_t^3
\left(1 - \frac{m_Z^2}{m_t^2}\right)
\left(2 - \frac{m_Z^2}{m_t^2} - \frac{m_Z^4}{m_t^4}\right).
\end{equation}

Using these expressions, the branching ratios are defined as
\begin{equation}
\mathrm{BR}(t \to qV) = \frac{\Gamma(t \to qV)}{\Gamma_t}, \quad (V = Z, \gamma).
\end{equation}

The above expressions correspond to the most general FCNC framework in which both left- and right-handed anomalous interactions are allowed. For completeness, these general formulae are retained throughout the manuscript. However, as demonstrated in Sec.~II, the production of right-handed top quarks is strongly suppressed at high center-of-mass energies owing to the chiral structure of the Standard Model $tbW$ vertex. Consequently, the detector-level simulation, statistical analysis, and the extraction of the projected limits presented in this work are performed exclusively for the left-handed anomalous couplings. Accordingly, when converting the projected coupling sensitivities into branching-ratio limits, the right-handed couplings are set to zero,
$\kappa_{qt}^{R}=\lambda_{qt}^{R}=0,$
such that all quoted branching-ratio limits correspond to purely left-handed FCNC interactions.

The corresponding upper limits on the FCNC branching ratios are summarized in Table~\ref{table:br}. The limits are obtained by translating the projected individual coupling sensitivities listed in  Table~\ref{tab:ss_limits} into branching fractions using Eqs.~(9)--(13), with only the left-handed anomalous  couplings retained in accordance with the detector-level analysis. The results
are presented separately for the real and imaginary components, with the values corresponding to the imaginary couplings given in square brackets.

For the $95\%$ C.L. exclusion scenario with $\delta_{\rm sys}=0\%$, branching-ratio sensitivities as low as
$\mathrm{BR}(t\to qZ)=6.77\times10^{-7}\,[6.20\times10^{-7}]$ and $\mathrm{BR}(t\to q\gamma)=4.97\times10^{-7}\,[3.66\times10^{-7}]$ are obtained for the real [imaginary] coupling components, respectively. At the $5\sigma$ discovery level, the corresponding sensitivities are $\mathrm{BR}(t\to qZ)=1.63\times10^{-6}\,[1.55\times10^{-6}]$ and $\mathrm{BR}(t\to q\gamma)=1.07\times10^{-6}\,[8.83\times10^{-7}]$. As expected, increasing the systematic uncertainty from $0\%$ to $10\%$ gradually weakens the achievable limits. Nevertheless, the projected sensitivities remain at the $\mathcal{O}(10^{-6})$ level, demonstrating the robustness of the proposed analysis strategy.

A comparison of the two decay channels shows that the $\mathrm{BR}(t\to q\gamma)$ channel generally provides stronger constraints than the $\mathrm{BR}(t\to qZ)$ channel. This behavior is consistent with the stronger sensitivity to the $\lambda_{qt}^{L}$ coupling observed in the corresponding two-dimensional contour analysis. Overall, the projected branching-ratio sensitivities demonstrate the strong potential of a high-energy muon collider for probing rare top-quark FCNC interactions.
 \begin{table}[httt!]
\caption{Projected upper limits on the branching ratios of the FCNC decays
$t\to qZ$ and $t\to q\gamma$, obtained by translating the individual
coupling sensitivities listed in Table~\ref{tab:ss_limits} using
Eqs.~(9)--(13). Values outside (inside) square brackets correspond to the
real (imaginary) components of the anomalous couplings.}
\centering
\begin{ruledtabular}
\begin{tabular}{c c c c }
 Coefficient & $\delta_{\rm sys}$  & $5\sigma$ & 95\% C.L. \\ [0.5ex]
 \hline\hline
    & $0\%$  & $1.63 \times 10^{-6}$ [$1.55 \times 10^{-6}$]  & $6.77 \times 10^{-7}$[$6.20 \times 10^{-7}$]  \\
 $BR(t\to qZ)$ & $5\%$ &   $2.69 \times 10^{-6}$[$2.65 \times 10^{-6}$]  & $1.06 \times 10^{-6}$[$1.03 \times 10^{-6}$] \\
  & $10\%$    & $4.83 \times 10^{-6}$[$4.93 \times 10^{-6}$]   & $1.60 \times 10^{-6}$[$1.63 \times 10^{-6}$]  \\
  \hline
  & $0\%$    & $1.07 \times 10^{-6}$[$8.83 \times 10^{-7}$]   & $4.97 \times 10^{-7}$[$3.66 \times 10^{-7}$]  \\
 $BR(t\to q\gamma)$ & $5\%$     & $1.74 \times 10^{-6}$[$1.46 \times 10^{-6}$]   & $8.64 \times 10^{-7}$[$6.62 \times 10^{-7}$]  \\
  & $10\%$ &  $3.14 \times 10^{-6}$ [$2.71 \times 10^{-6}$]  & $1.49 \times 10^{-7}$[$1.15 \times 10^{-6}$]  \\
\end{tabular}
\end{ruledtabular}
\label{table:br}
\end{table}

\begin{table*}[htbp]
\centering
\caption{Comparison of the projected $95\%$ C.L. upper limits on the FCNC
top-quark branching ratios $\mathrm{BR}(t\rightarrow qZ)$ and
$\mathrm{BR}(t\rightarrow q\gamma)$ obtained at current hadron colliders,
representative future $e^+e^-$ collider proposals, and the $10$ TeV muon
collider considered in this work. The limits quoted for this work correspond
to the individual coupling sensitivities for the real components of the
anomalous couplings in the $\delta_{\rm sys}=0\%$ scenario, as obtained from
the two-dimensional contour analysis.}
\label{tab:lepton_colliders_comp}
\begin{tabular}{lcccccc}
\hline\hline
Collider & $\sqrt{s}$ (TeV) & $L_{\rm int}$ (ab$^{-1}$) & Process & $\mathrm{BR}(t\rightarrow qZ)$ & $\mathrm{BR}(t\rightarrow q\gamma)$ & Reference \\
\hline
ATLAS coll.    & 13 & 0.139 & $pp \to t\bar t +t\gamma(Z)$           & $6.20 \times 10^{-5}$& $0.85 \times 10^{-5}$ & \cite{ATLAS:2022per,ATLAS:2023qzr} \\
CMS coll.    & 13 & 0.138 & $pp \to t\bar t +t\gamma(Z)$           & $2.20 \times 10^{-4}$& $0.95 \times 10^{-5}$ & \cite{CMS:2017wcz,CMS:2023bjm}
\\
FCC-ee    & 0.350 & 10  & $e^+e^- \to t\bar{q}$           & $7.60 \times 10^{-6}$& $5.25 \times 10^{-6}$ & \cite{Khanpour:2014xla} \\
CLIC      & 0.380  & 1.0 & $e^+e^- \to t\bar{q}$           & --- & $2.60 \times 10^{-5}$ & \cite{CLICdp:2018esa} \\
Future LC & 0.350  & 3.0 & $e^+e^- \to t\bar{q}Z / \gamma$ & $4.5 \times 10^{-4}$ & $1.7 \times 10^{-4}$ &  \cite{Khatibi:2021phr} \\
\hline
Muon Collider & 10 & 10 & $\mu^+\mu^-\rightarrow\nu_\mu\mu^+bj$ 
& \textbf{$6.77\times10^{-7}$} 
& \textbf{$4.97\times10^{-7}$} 
& This work \\
\hline\hline
\end{tabular}
\end{table*}

Table~\ref{tab:lepton_colliders_comp} summarizes the projected $95\%$ C.L. sensitivities obtained in this work together with the current experimental limits reported by the ATLAS and CMS Collaborations and the projected reaches of representative future lepton collider studies. The comparison clearly illustrates the substantial improvement expected from a high-energy muon collider operating at $\sqrt{s}=10$~TeV with an integrated luminosity of $10~\mathrm{ab}^{-1}$.

Compared with the present LHC constraints, the projected sensitivities improve significantly in both FCNC decay channels. For the $t\rightarrow qZ$ decay, the expected upper limit reaches $6.77\times10^{-7}$, improving upon the current ATLAS and CMS limits by approximately two and three orders of magnitude, respectively. An even more pronounced improvement is obtained for the $t\rightarrow q\gamma$ channel, where the projected limit of $4.97\times10^{-7}$ exceeds the current experimental bounds by more than one order of magnitude. These results demonstrate the remarkable discovery potential of a multi-TeV muon collider for rare top-quark FCNC interactions.

The comparison with future electron--positron collider proposals is equally encouraging. While FCC-ee is expected to achieve excellent sensitivity owing to its clean experimental environment, its lower center-of-mass energy limits the kinematic reach for single-top FCNC production. Consequently, the projected sensitivity obtained in the present study is substantially stronger for both branching ratios. A similar conclusion can be drawn with respect to the CLIC and other proposed linear collider analyses, whose projected limits remain at the level of $10^{-5}$--$10^{-4}$ for most channels. In contrast, the present analysis reaches the $\mathcal{O}(10^{-7})$ sensitivity for both $t\rightarrow qZ$ and $t\rightarrow q\gamma$ decays.

The superior sensitivity primarily originates from the unique combination of the high center-of-mass energy ($10$~TeV), the large integrated luminosity ($10~\mathrm{ab}^{-1}$), and the clean experimental environment provided by a muon collider. In addition, the use of a multivariate BDT analysis substantially enhances the signal-to-background discrimination, allowing rare FCNC processes to be probed with unprecedented precision. These results highlight the complementary role of future muon colliders within the global program of precision top-quark physics and establish them as one of the most promising facilities for exploring flavor-changing neutral current interactions beyond the Standard Model.
\section{CONCLUSION}\label{section6}

In this work, we have performed a comprehensive investigation of flavor-changing neutral current (FCNC) interactions involving the top quark through the anomalous $tqZ$ and $tq\gamma$ vertices at a future muon collider operating at a center-of-mass energy of $\sqrt{s}=10$~TeV with an integrated luminosity of $10~\mathrm{ab}^{-1}$. The study has been carried out within the framework of the most general effective FCNC Lagrangian, including both vector and tensor interactions with independent left- and right-handed chiral structures.

As a first step, the relative contributions of the different effective operators were examined at the parton level. Owing to the chiral structure of the Standard Model $tbW$ interaction, right-handed tensor couplings were found to be strongly suppressed at multi-TeV energies, while the vector interaction $X_{qt}$ gives a considerably smaller contribution than the tensor operators over the parameter space considered. Motivated by these observations, the detector-level analysis was performed by focusing on the left-handed tensor couplings $\kappa_{qt}^{L}$ and $\lambda_{qt}^{L}$, which provide the dominant sensitivity to FCNC interactions in the process under investigation.

A realistic simulation chain has been employed throughout the analysis, including the implementation of the effective Lagrangian in {\sc FeynRules}, event generation with {\sc MadGraph5\_aMC@NLO}, parton showering and hadronization with {\sc Pythia8}, detector simulation using {\sc Delphes}, and event reconstruction with {\sc ROOT}. To maximize the separation between signal and the relevant Standard Model backgrounds, a multivariate analysis based on Boosted Decision Trees (BDTs) was developed. In addition to conventional kinematic observables, angular variables sensitive to the interference between the $tqZ$ and $tq\gamma$ interactions were incorporated into the BDT training, resulting in a substantial enhancement of the signal discrimination.

Unlike conventional one-dimensional analyses in which a single anomalous coupling is varied at a time, the present study performs a simultaneous two-parameter analysis by allowing both $\kappa_{qt}^{L}$ and $\lambda_{qt}^{L}$ to vary simultaneously. The projected sensitivities are presented as two-dimensional exclusion and discovery contours in both the $\mathrm{Re}(\kappa_{qt}^{L})$--$\mathrm{Re}(\lambda_{qt}^{L})$ and $\mathrm{Im}(\kappa_{qt}^{L})$--$\mathrm{Im}(\lambda_{qt}^{L})$ parameter planes. This approach naturally incorporates the interplay between the two FCNC interactions and provides a more realistic determination of the accessible parameter space.

The obtained results demonstrate that anomalous FCNC couplings can be probed at the level of a few $10^{-3}$, corresponding to branching-ratio sensitivities of the order of $\mathcal{O}(10^{-6})$ for the rare decays $t\rightarrow qZ$ and $t\rightarrow q\gamma$. The projected limits remain remarkably stable even after including systematic uncertainties up to $\delta_{\rm sys}=10\%$, indicating the robustness of the proposed analysis strategy. Furthermore, the sensitivities obtained for the real and imaginary components of the anomalous couplings are found to be very similar, showing that the collider reach is primarily determined by the magnitude of the effective interactions.

The corresponding branching-ratio limits significantly improve upon the current experimental bounds reported by the ATLAS and CMS Collaborations and are competitive with, or superior to, the projected sensitivities of other proposed future lepton collider facilities. In particular, the exceptional center-of-mass energy available at a multi-TeV muon collider substantially enhances the sensitivity to dipole-type FCNC operators, making this machine an especially powerful probe of rare top-quark interactions.

Finally, it is worth emphasizing that the projected sensitivities remain several orders of magnitude above the extremely suppressed Standard Model expectations, $\mathrm{BR}(t\rightarrow qZ)\sim10^{-14}$ and $\mathrm{BR}(t\rightarrow q\gamma)\sim10^{-12}$. Consequently, any observation within the parameter space explored in this work would constitute a clear indication of physics beyond the Standard Model. Overall, the process
\[
\mu^{+}\mu^{-}\rightarrow\nu_{\mu}\,\mu^{+}\,b\,j
\]
emerges as one of the most promising channels for investigating anomalous top-quark FCNC interactions at future high-energy muon colliders, providing an important and complementary avenue in the global program of precision top-quark physics and searches for new phenomena.
\begin{acknowledgments}
 The numerical calculations reported in this paper were partially performed at TUBITAK ULAKBIM, High Performance and Grid Computing Center (TRUBA resources).
\end{acknowledgments}


\begin{thebibliography}{99}
\bibitem{Glashow70} S.~L.~Glashow, J.~Iliopoulos, and L.Maiani, Phys. Rev. D \textbf{2}, 1285 (1970).

\bibitem{Eilam:1990zc} 
  G.~Eilam, J.~L.~Hewett and A.~Soni,
  Phys.\ Rev.\ D {\bf 44}, 1473 (1991), Phys.\ Rev.\ D {\bf 59}, 039901(E) (1998)].
  
\bibitem{Yang:1997dk} 
  J.~M.~Yang, B.~L.~Young and X.~Zhang,
  Phys.\ Rev.\ D {\bf 58}, 055001 (1998)
  [hep-ph/9705341].
  
\bibitem{Lu:2003yr} 
  G.~r.~Lu, F.~r.~Yin, X.~l.~Wang and L.~d.~Wan,
  Phys.\ Rev.\ D {\bf 68}, 015002 (2003)
  [hep-ph/0303122].

\bibitem{Li:1993mg} 
  C.~S.~Li, R.~J.~Oakes and J.~M.~Yang,
  Phys.\ Rev.\ D {\bf 49}, 293 (1994)
  Erratum: [Phys.\ Rev.\ D {\bf 56}, 3156 (1997)].
  
 \bibitem{AguilarSaavedra:2004wm}
J.~A.~Aguilar-Saavedra,
Acta Phys.\ Polon.\ B \textbf{35} (2004), 2695
[arXiv:hep-ph/0409342 [hep-ph]].
 
 \bibitem{AguilarSaavedra:2008zc}
J.~A.~Aguilar-Saavedra,
Nucl.\ Phys.\ B \textbf{812} (2009), 181--204
[arXiv:0811.3842 [hep-ph]].

\bibitem{Aguilar-Saavedra:2009ygx}
J.~A.~Aguilar-Saavedra,
Nucl. Phys. B \textbf{821} (2009), 215-227
[arXiv:0904.2387 [hep-ph]].

\bibitem{OPAL:2001spi}
G.~Abbiendi \textit{et al.} [OPAL],
Phys. Lett. B \textbf{521}, 181-194 (2001)
[arXiv:hep-ex/0110009 [hep-ex]].

\bibitem{ALEPH:2002wad}
A.~Heister \textit{et al.} [ALEPH],
Phys. Lett. B \textbf{543}, 173-182 (2002)
[arXiv:hep-ex/0206070 [hep-ex]].

\bibitem{L3:2002hbp}
P.~Achard \textit{et al.} [L3],
Phys. Lett. B \textbf{549}, 290-300 (2002)
[arXiv:hep-ex/0210041 [hep-ex]].

\bibitem{DELPHI:2003cnx}
J.~Abdallah \textit{et al.} [DELPHI],
Phys. Lett. B \textbf{590}, 21-34 (2004)
[arXiv:hep-ex/0404014 [hep-ex]].

\bibitem{ZEUS:2011mya}
H.~Abramowicz \textit{et al.} [ZEUS],
Phys. Lett. B \textbf{708}, 27-36 (2012)
[arXiv:1111.3901 [hep-ex]].

\bibitem{CDF:2008mpz}
T.~Aaltonen \textit{et al.} [CDF],
Phys. Rev. Lett. \textbf{101}, 192002 (2008)
[arXiv:0805.2109 [hep-ex]].

\bibitem{D0:2011pcl}
V.~M.~Abazov \textit{et al.} [D0],
Phys. Lett. B \textbf{701}, 313-320 (2011)
[arXiv:1103.4574 [hep-ex]].

\bibitem{ATLAS:2015vhj}
G.~Aad \textit{et al.} [ATLAS],
Eur. Phys. J. C \textbf{76}, no.1, 12 (2016)
[arXiv:1508.05796 [hep-ex]].

\bibitem{ATLAS:2018zsq}
M.~Aaboud \textit{et al.} [ATLAS],
JHEP \textbf{07}, 176 (2018)
[arXiv:1803.09923 [hep-ex]].

\bibitem{ATLAS:2022per}
G.~Aad \textit{et al.} [ATLAS],
Phys. Lett. B \textbf{842} (2023), 137379
[erratum: Phys. Lett. B \textbf{847} (2024), 138286]
[arXiv:2205.02537 [hep-ex]].

\bibitem{CMS:2013knb}
S.~Chatrchyan \textit{et al.} [CMS],
Phys. Rev. Lett. \textbf{112}, no.17, 171802 (2014)
[arXiv:1312.4194 [hep-ex]].

\bibitem{CMS:2017wcz}
A.~M.~Sirunyan \textit{et al.} [CMS],
JHEP \textbf{07}, 003 (2017)
[arXiv:1702.01404 [hep-ex]].

\bibitem{ATLAS:2023qzr}
G.~Aad \textit{et al.} [ATLAS],
Phys. Rev. D \textbf{108}, no.3, 032019 (2023)
[arXiv:2301.11605 [hep-ex]].

\bibitem{CMS:2023bjm}
A.~Hayrapetyan \textit{et al.} [CMS],
Phys. Rev. D \textbf{109} (2024) no.7, 072004
[arXiv:2312.08229 [hep-ex]].

\bibitem{Han:1998yr}
T.~Han and J.~L.~Hewett,
Phys. Rev. D \textbf{60} (1999), 074015
[arXiv:hep-ph/9811237 [hep-ph]].

\bibitem{Yang:2004af}
J.~M.~Yang,
Annals Phys. \textbf{316} (2005), 529-539
[arXiv:hep-ph/0409351 [hep-ph]].

\bibitem{Khanpour:2014xla}
H.~Khanpour, S.~Khatibi, M.~Khatiri Yanehsari and M.~Mohammadi Najafabadi,
Phys. Lett. B \textbf{775} (2017), 25-31
[arXiv:1408.2090 [hep-ph]].

\bibitem{Tizchang:2024ctw}
S.~Tizchang, H.~Khanpour and M.~Mohammadi Najafabadi,
Iran. J. Phys. Res. \textbf{24} (2024) no.4, 333-341

\bibitem{Khatibi:2021phr}
S.~Khatibi and M.~Moallemi,
J. Phys. G \textbf{48} (2021) no.12, 125004
[arXiv:2106.08231 [hep-ph]].

\bibitem{Shi:2019epw}
L.~Shi and C.~Zhang,
Chin. Phys. C \textbf{43} (2019) no.11, 113104
[arXiv:1906.04573 [hep-ph]].

\bibitem{CLICdp:2018esa}
H.~Abramowicz \textit{et al.} [CLICdp],
JHEP \textbf{11} (2019), 003
[arXiv:1807.02441 [hep-ex]].

\bibitem{Cakir:2018ruj}
O.~Cakir, A.~Yilmaz, I.~Turk Cakir, A.~Senol and H.~Denizli,
Nucl. Phys. B \textbf{944} (2019), 114640
[arXiv:1809.01923 [hep-ph]].

\bibitem{TurkCakir:2017rvu}
I.~Turk Cakir, A.~Yilmaz, H.~Denizli, A.~Senol, H.~Karadeniz and O.~Cakir,
Adv. High Energy Phys. \textbf{2017} (2017), 1572053
[arXiv:1705.05419 [hep-ph]].

\bibitem{Denizli:2017cfx}
H.~Denizli, A.~Senol, A.~Yilmaz, I.~Turk Cakir, H.~Karadeniz and O.~Cakir,
Phys. Rev. D \textbf{96} (2017) no.1, 015024
[arXiv:1701.06932 [hep-ph]].

\bibitem{mu1}
K.~Long, D.~Lucchesi, M.~Palmer, N.~Pastrone, D.~Schulte and V.~Shiltsev,
Nature Phys. \textbf{17}, no.3, 289-292 (2021)
[arXiv:2007.15684 [physics.acc-ph]].

\bibitem{mu2} 
C.~Accettura, D.~Adams, R.~Agarwal, C.~Ahdida, C.~Aim\`e, N.~Amapane, D.~Amorim, P.~Andreetto, F.~Anulli and R.~Appleby, \textit{et al.}
Eur. Phys. J. C \textbf{83}, no.9, 864 (2023)
[erratum: Eur. Phys. J. C \textbf{84}, no.1, 36 (2024)]
[arXiv:2303.08533 [physics.acc-ph]].

\bibitem{mu3}
D.~Stratakis \textit{et al.} [Muon Collider],
[arXiv:2203.08033 [physics.acc-ph]].

\bibitem{mu6}
Y.~F.~Dong, Y.~C.~Mao, i.~C.~Yang and J.~C.~Yang,
Eur. Phys. J. C \textbf{83}, no.7, 555 (2023)
[arXiv:2304.01505 [hep-ph]].

\bibitem{mu7}
T.~Li, C.~Y.~Yao and M.~Yuan,
JHEP \textbf{09}, 131 (2023)
[arXiv:2306.17368 [hep-ph]].


\bibitem{Accettura:2023ked}
C.~Accettura, D.~Adams, R.~Agarwal, C.~Ahdida, C.~Aim{\`e}, N.~Amapane, D.~Amorim, P.~Andreetto, F.~Anulli and R.~Appleby, \textit{et al.}
Eur. Phys. J. C \textbf{83} (2023) no.9, 864
[erratum: Eur. Phys. J. C \textbf{84} (2024) no.1, 36]
[arXiv:2303.08533 [physics.acc-ph]].

\bibitem{Ake:2023xcz}
D.~Ake, A.~O.~Bouzas and F.~Larios,
Adv. High Energy Phys. \textbf{2024} (2024), 2038180
[arXiv:2311.09488 [hep-ph]].

\bibitem{Han:2024gan}
T.~Han, D.~Liu and S.~Wang,
Phys. Rev. D \textbf{111} (2025) no.3, 035015
[arXiv:2410.11015 [hep-ph]].

\bibitem{Aguilar-Saavedra:2010uur}
J.~A.~Aguilar-Saavedra,
Nucl. Phys. B \textbf{843} (2011), 638-672
[erratum: Nucl. Phys. B \textbf{851} (2011), 443-444]
[arXiv:1008.3562 [hep-ph]].

\bibitem{Grzadkowski:2010es}
B.~Grzadkowski, M.~Iskrzynski, M.~Misiak and J.~Rosiek,
JHEP \textbf{10} (2010), 085
[arXiv:1008.4884 [hep-ph]].

\bibitem{Aguilar-Saavedra:2018ksv}
J.~A.~Aguilar-Saavedra, C.~Degrande, G.~Durieux, F.~Maltoni, E.~Vryonidou, C.~Zhang, D.~Barducci, I.~Brivio, V.~Cirigliano and W.~Dekens, \textit{et al.}
[arXiv:1802.07237 [hep-ph]].

\bibitem{Degrande:2011ua}
C.~Degrande, C.~Duhr, B.~Fuks, D.~Grellscheid, O.~Mattelaer and T.~Reiter,
Comput. Phys. Commun. \textbf{183} (2012), 1201-1214
[arXiv:1108.2040 [hep-ph]].

\bibitem{Alloul:2013bka}
A.~Alloul, N.~D.~Christensen, C.~Degrande, C.~Duhr and B.~Fuks,
Comput. Phys. Commun. \textbf{185} (2014), 2250-2300
[arXiv:1310.1921 [hep-ph]].

\bibitem{Alwall:2014hca}
J.~Alwall, R.~Frederix, S.~Frixione, V.~Hirschi, F.~Maltoni, O.~Mattelaer, H.~S.~Shao, T.~Stelzer, P.~Torrielli and M.~Zaro,
JHEP \textbf{07} (2014), 079
[arXiv:1405.0301 [hep-ph]].

\bibitem{Sjostrand:2014zea}
T.~Sj\"ostrand, S.~Ask, J.~R.~Christiansen, R.~Corke, N.~Desai, P.~Ilten, S.~Mrenna, S.~Prestel, C.~O.~Rasmussen and P.~Z.~Skands,
Comput. Phys. Commun. \textbf{191} (2015), 159-177
[arXiv:1410.3012 [hep-ph]].

\bibitem{deFavereau:2013fsa}
J.~de Favereau \textit{et al.} [DELPHES 3],
JHEP \textbf{02} (2014), 057
[arXiv:1307.6346 [hep-ex]].

\bibitem{Cacciari:2011ma} 
  M.~Cacciari, G.~P.~Salam and G.~Soyez,
  Eur.\ Phys.\ J.\ C {\bf 72}, 1896 (2012)[arXiv:1111.6097 [hep-ph]].

  \bibitem{Cacciari:2008gp} 
  M.~Cacciari, G.~P.~Salam and G.~Soyez,
  JHEP {\bf 0804}, 063 (2008)
[arXiv:0802.1189 [hep-ph]].

\bibitem{Cowan:2010js}
G.~Cowan, K.~Cranmer, E.~Gross and O.~Vitells,
Eur. Phys. J. C \textbf{71}, 1554 (2011)
[erratum: Eur. Phys. J. C \textbf{73}, 2501 (2013)]
[arXiv:1007.1727 [physics.data-an]].

\bibitem{Kumar:2015tna}
N.~Kumar and S.~P.~Martin,
Phys. Rev. D \textbf{92}, no.11, 115018 (2015)
[arXiv:1510.03456 [hep-ph]].

\end{thebibliography}
\end{document}